\documentclass[twocolumn]{aastex63}

\usepackage{natbib}
\usepackage[]{threeparttable}
\usepackage[caption = false]{subfig}
\usepackage{graphicx}

\usepackage{rotating}       
\usepackage{mathrsfs}
\usepackage{array}

\usepackage{color}
\usepackage{amsmath}

\usepackage{multirow,tabularx}
\usepackage{calc}

\newsavebox{\twofigures}

\usepackage{silence}
\graphicspath{{./}{figures/}}

\shorttitle{A comprehensive catalog of NGC~5128 GC candidates}
\shortauthors{Hughes et al.}

\begin{document}

\title{NGC~5128 globular cluster candidates out to 150 kpc: a comprehensive catalog from {\it Gaia} and ground based data
\footnote{This paper includes data gathered with the 6.5 m Magellan Telescope at Las Campanas Observatory, Chile.}}

\correspondingauthor{Allison K. Hughes}
\email{akhughes@email.arizona.edu}

\author[0000-0002-1718-0402]{Allison K. Hughes}
\affil{Steward Observatory, University of Arizona, 933 North Cherry Avenue, Tucson, AZ 85721, USA}

\author[0000-0003-4102-380X]{David J. Sand}
\affil{Steward Observatory, University of Arizona, 933 North Cherry Avenue, Tucson, AZ 85721, USA}

\author[0000-0003-0248-5470]{Anil Seth}
\affiliation{Department of Physics \& Astronomy, University of Utah, Salt Lake City, UT, 84112, USA}

\author[0000-0002-1468-9668]{Jay Strader}
\affiliation{Center for Data Intensive and Time Domain Astronomy, Department of Physics and Astronomy, Michigan State University, East Lansing, MI 48824, USA}

\author[0000-0001-6215-0950]{Karina Voggel}
\affiliation{Universite de Strasbourg, CNRS, Observatoire astronomique de Strasbourg, UMR 7550, F-67000 Strasbourg, France}

\author[0000-0003-0234-3376]{Antoine Dumont}
\affiliation{Department of Physics \& Astronomy, University of Utah, Salt Lake City, UT, 84112, USA}

\author[0000-0002-1763-4128]{Denija Crnojevi\'{c}}
\affiliation{University of Tampa, 401 West Kennedy Boulevard, Tampa, FL 33606, USA}


\author[0000-0003-2352-3202]{Nelson Caldwell}
\affiliation{Center for Astrophysics, Harvard \& Smithsonian, 60 Garden Street, Cambridge, MA 02138, USA}

\author{Duncan A. Forbes}
\affiliation{Centre for Astrophysics and Supercomputing, Swinburne University of Technology, Hawthorn VIC 3122, Australia}

\author{Joshua D. Simon}
\affiliation{Observatories of the Carnegie Institution for Science, 813 Santa Barbara Street, Pasadena, CA 91101, USA}

\author[0000-0001-8867-4234]{Puragra Guhathakurta}
\affiliation{UCO/Lick Observatory, University of California Santa Cruz, 1156 High Street, Santa Cruz, CA 95064, USA}

\author[0000-0001-6443-5570]{Elisa Toloba}
\affiliation{Department of Physics, University of the Pacific, 3601 Pacific Avenue, Stockton, CA 95211, USA}

\begin{abstract}

We present a new catalog of 40502 globular cluster (GC) candidates in NGC~5128 out to a projected radius of $\sim$150 kpc, based on data from the Panoramic Imaging Survey of Centaurus and Sculptor (PISCeS), {\it Gaia} Data Release 2, and the NOAO Source Catalog. 
Ranking these candidates based on the likelihood that they are true GCs, we find that approximately 1900 belong to our top two ranking categories and should be the highest priority for spectroscopic follow-up for confirmation. Taking into account our new data and a vetting of previous GC catalogs, we estimate a total GC population of $1450 \pm 160$ GCs. We show that a substantial number of sources previously argued to be low-velocity GCs are instead foreground stars, reducing the inferred GC velocity dispersion. This work showcases the power of {\it Gaia} to identify slightly extended sources at the $\sim 4$ Mpc distance of NGC~5128, enabling accurate identification of GCs throughout the entire extended halo, not just the inner regions that have been the focus of most previous work. 


\end{abstract}
 


\section{Introduction}

Globular clusters (GCs) are an essential tool for understanding the formation, structure, and evolution of galaxies beyond the Local Group, where spectroscopy of individual stars is typically difficult or impossible (see the reviews of \citealt{Harris91,Brodie06,Beasley20}). These old, massive ($\gtrsim 10^4 M_{\odot}$) star clusters are found in all but the least luminous galaxies. The majority of GCs are associated with the stellar halo, where they serve as luminous beacons of the underlying low surface brightness stellar population. With modern wide-field imagers and spectrographs, GCs can be used to measure the mass and structure of dark matter halos (e.g., \citealt{Alabi16,Wasserman18}) and trace individual accretion events (e.g., \citealt{Mackey10}).

Historically, NGC~5128 (Centaurus A) has been a leading target for extragalactic GC studies, due to its richness and proximity.  NGC~5128 is the central elliptical galaxy in a small group of galaxies at a distance of $3.8 \pm 0.1$ Mpc \citep{Harris2010}. Since the first discovery of GCs in NGC~5128 in the 1980s, many photometric and spectroscopic surveys have been conducted, leading to the identification of 557 confirmed GCs and thousands of GC candidates \citep{Graham1980, VanDenBergh1981, Hesser1986, Harris1992,  Harris2004,  Harris2012, Holland1999, Harris2002, Harris2006, Peng2004, Martini2004,   Woodley2005, Woodley2007, Woodley2010a, Woodley2010b, Gomez2006,  Rejkuba2007, Beasley2008, Georgiev2009,  Georgiev2010, Taylor2010, Taylor2015, Taylor2017, Mouhcine2010, Sinnott2010,  Voggel2020, Fahrion2020, Muller2020}. 
The majority of these studies have focused on the inner $\sim$~40 kpc of the galaxy. Both the presence of halo substructures out to $\gtrsim$~100 kpc \citep{crnojevic16,Crnojevic19} and the fact that the total GC population of NGC~5128 is likely in the range of $\sim$~1000--2000 clusters (e.g., \citealt{Harris2006,Harris2010_alone, Taylor2017}) illustrate that there is substantial room for novel studies of the NGC~5128 GC system, especially in the outer halo.

While the proximity of NGC~5128 aids in the identification and follow-up of GC candidates,
this close distance also provides several challenges. First, the halo of NGC~5128 is spread across $\gtrsim$ 20 deg$^2$ of sky \citep{crnojevic16}, requiring large area searches for a comprehensive accounting of the stellar halo and GC system. NGC~5128 is also relatively close to the Galactic plane (with a Galactic latitude of $b \sim 19^{\circ}$), leading to significant contamination from foreground stars as well as foreground extinction ($E(B-V) \sim 0.12$ mag; \citealt{Schlafly2011}). An additional issue is that with a modest systematic velocity of 
541 km s$^{-1}$ \citep{Hui1995} and a velocity dispersion of $\sigma \sim 150$ km s$^{-1}$ \citep{Wilkinson1986,Silge05}, there is some overlap in radial velocity of NGC~5128 GCs and Milky Way  foreground stars: radial velocity \emph{alone} cannot definitively give membership for every object, especially on the low-velocity end of the distribution (see Appendix \ref{app:misconfirmed}).

The Panoramic Imaging Survey of Centaurus and Sculptor (PISCeS) produced a wide-field resolved stellar map of NGC~5128 out to a projected galactocentric radius of $\sim$ 150 kpc \citep{crnojevic16}. The PISCeS data are invaluable in searching for GCs because of the survey's wide angular coverage and good seeing (median FWHM of $0.66\arcsec$), which allows GCs at the distance of NGC~5128 to be marginally resolved and hence separated from foreground stars. 

Here we take advantage of this exquisite dataset to build a new wide-field catalog of GC candidates, focusing especially on the outer halo (projected radii $\sim 40$--150 kpc) that is nearly unexplored with spectroscopy. We supplement our high-quality PISCeS imaging with other data from {\it Gaia} DR2 and the NOAO Source Catalog to refine our sample, with a future goal of comprehensive spectroscopic follow-up.

The PISCeS data are well-suited to identify typical GCs outside the central regions of NGC~5128. In a companion paper \citep{Voggel2020}, hereafter V20, our team focuses on the
related question of constructing a complete sample of the most luminous GCs and stripped galaxy nuclei in NGC~5128. This requires
a distinct approach as many of these sources are saturated or crowded in the PISCeS data.

This paper is organized as follows.  Section~\ref{sec:data} outlines the key datasets used in our GC candidate search.  In Section~\ref{sec:test sample} we identify a fiducial sample of known radial velocity confirmed NGC~5128 GCs which we use to define our GC candidate selection methodology, which we describe in Section~\ref{sec:GCselection}.  Our final catalogs are presented in Section~\ref{sec:catsec}, and further results are discussed in Section~\ref{sec:discussion}.  We summarize and conclude in Section~\ref{sec:conclude}.  In the Appendix, we present a collected catalog of all confirmed NGC~5128 GCs as a resource for the community.  We additionally provide information detailing foreground stars, identified by their {\it Gaia} parallax or proper motion measurements, that were previously mis-reported as NGC~5128 GCs, and report and resolve other discrepancies in the NGC~5128 GC literature, when possible. Throughout this work we adopt a distance modulus for NGC~5128 of
$(m - M)_{0} = 27.91$ mag, corresponding to a distance of $D =3.82$ Mpc \citep{Harris2010}. The physical scale at this distance is 18.5 pc~arcsec$^{-1}$ (1.1 kpc~arcmin$^{-1}$).


\section{Key Datasets} \label{sec:data}

\subsection{The PISCeS Survey} \label{sub:pisces}

The foundational dataset of our GC search is deep NGC~5128 Magellan/Megacam imaging from the PISCeS survey.  Here we briefly describe the relevant aspects of the survey strategy and observational methods; for further details see \citet{Crnojevic14,crnojevic16,Crnojevic19}.

The goal of the PISCeS survey is to image the halos of the nearby massive galaxies NGC~5128 and NGC~253
(NGC~253 is not covered in this paper; see \citealt{Sand14} and \citealt{Toloba16}) out to a projected radius of $R \sim$ 150 kpc. The data are deep enough to reach $\sim$ 1--2 mag below the tip of the red giant branch ($r \sim 25.5$--26.5 mag). These uniquely deep wide-field maps enable discovery of faint dwarf galaxies and other halo substructures, such as tidal streams and shells, allowing comparisons with similar observations in the Local Group \citep[e.g.][]{McConnachie09}.

\begin{figure}[t]
\includegraphics[width=8.5cm]{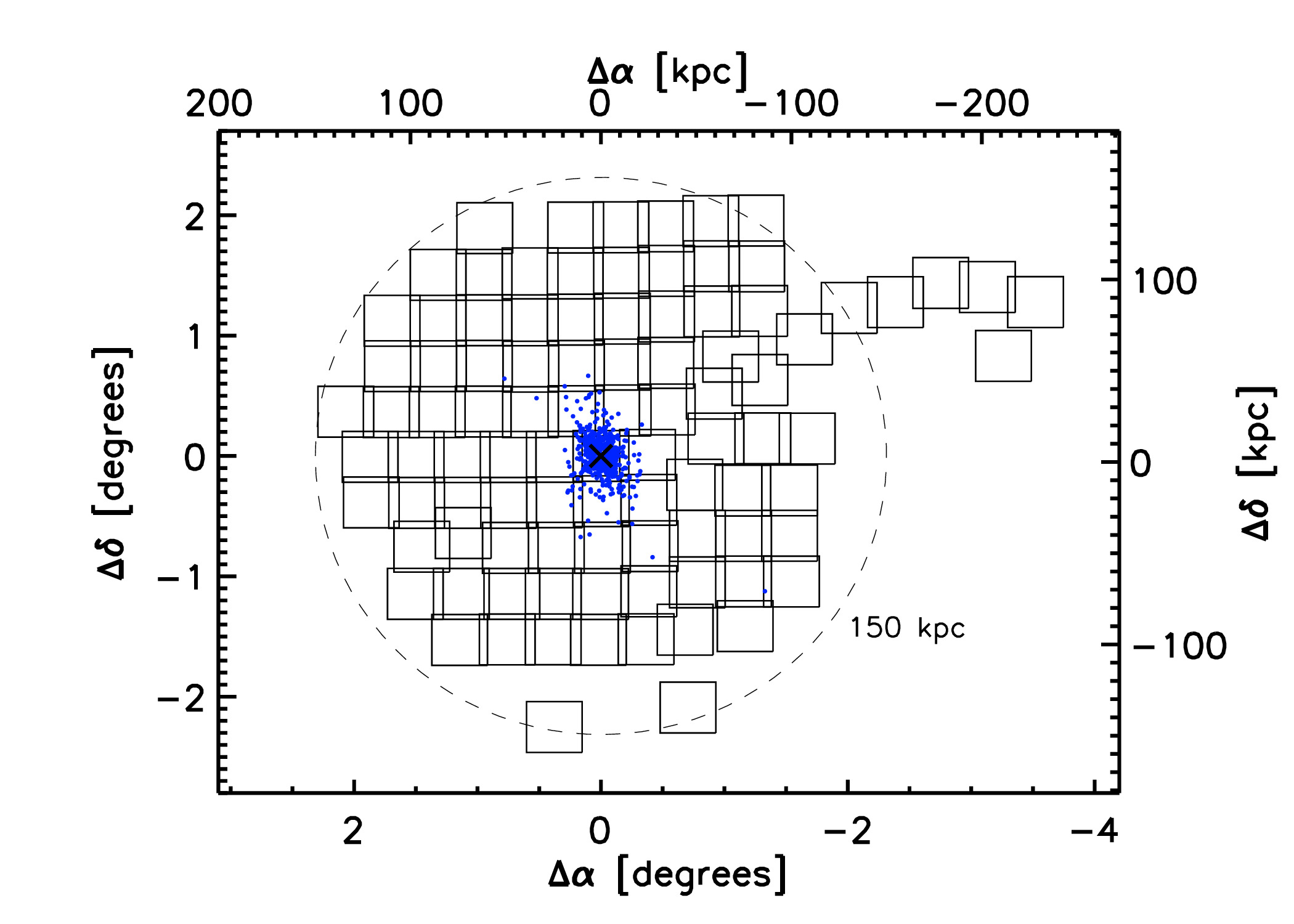}
\caption{Footprint of the PISCeS survey around NGC~5128, using data from 2010-2017 and  comprising 95 separate Magellan/Megacam fields and oriented such that north is up and east is left.  The dashed circle marks a radius of 150 kpc from the center of NGC~5128, located at $\alpha$=$201.362540^{\circ}$, $\delta$=$-43.033627^{\circ}$, which is marked by the central black `x'. The blue dots mark the positions of known confirmed GCs around NGC~5128, which are largely confined to the central $\sim$40 kpc.  Note that several pointings in this map were not presented in the dwarf galaxy luminosity function analysis of \citet{Crnojevic19} but  are included here: an extension to the northwest, two pointings to the south centered on known dwarfs KK203 and KK196, and data that was not high enough quality for a dwarf search but is sufficiently deep for our GC search. }
\label{fig:map}
\end{figure}

The PISCeS survey uses the Megacam imager \citep{McLeod2015} on the 6.5-m Magellan Clay telescope. With the f/5 secondary, Megacam has a $\sim$ 24\arcmin\ $\times$ 24\arcmin\ field-of-view with a binned (2$\times$2) scale of 0.16\arcsec\ per pixel. We tiled the halo of NGC~5128 with individual pointings of this size, as illustrated in Figure~\ref{fig:map}. The typical exposure times were 6$\times$300~s in both the $g$ and $r$ bands, but were adjusted depending on the seeing and sky conditions to try to achieve a similar  depth in poorer seeing or non-photometric conditions.

The data were reduced in the standard manner at the Smithsonian Astrophysical Observatory Telescope Data Center, including bias subtraction, flat fielding, and cosmic ray rejection. 
As part of this reduction, an ICRS astrometric solution based on the 2MASS point source catalog \citep{Cutri03} was applied and the images of a given field and filter were stacked with {\tt SWARP} \citep{Bertin2010}. Subsequent to this processing, the release of {\it Gaia} DR2 allowed us to check this astrometry. Cross-matching found a mean PISCeS--{\it Gaia} DR2 offset of ($\alpha$ cos $\delta$, $\delta$) of ($0.1300\pm0.0002\arcsec$, $0.0634\pm0.0002\arcsec$). These values are consistent across fields and subsamples of PISCeS. We do \emph{not} correct the self-consistent astrometry of PISCeS to the {\it Gaia} DR2 frame.

Photometric calibration was done by observing Sloan Digital Sky Survey (SDSS; \citealt{Ahumada20}) equatorial fields on photometric nights at a variety of airmasses. 
We designed intentional $\sim 2\arcmin$ overlaps between NGC~5128 fields to allow for cross-calibration of the entire survey, leveraging data taken on photometric nights. All photometry is on the AB mag system.
This paper includes data from 95 pointings around NGC~5128 obtained between 2010 and 2017, as illustrated in Figure~\ref{fig:map}.  The footprint includes nearly full coverage of the area out to $\sim$ 150 kpc from the center of the galaxy, and an arm extending in the north-west direction out to $\sim 260$ kpc.

For GC candidate photometry, we ran {\tt Source Extractor} \citep{Bertin96} on the $g$ and $r$ stacked images. Aperture photometry was done using aperture diameters of 3 and 6 pixels (0.48\arcsec and 0.96\arcsec, respectively), with the intention of using the differences between these magnitudes to select extended sources---including GCs.   
We also measure the ``total" magnitudes for each source in our catalog at a diameter of 6$\times$FWHM (for the median seeing, this is a radius $\sim 12.4$ pixels = 2.0\arcsec ), with an aperture correction determined on a field-by-field basis. When relevant, we correct the photometry for foreground Galactic extinction using the maps of \citet{Schlafly2011} on a source-by-source basis. We are explicit about when a correction for extinction is (or is not) made throughout this work, and all absolute magnitudes are extinction-corrected. Finally, we combine the $g$ and $r$ catalogs using a matching radius of $1\arcsec$. Our use of PISCeS data to select GC candidates is detailed in \S 4.1--4.3.

GCs are luminous: the peak of the lognormal GC luminosity function (GCLF) is expected to occur at $M_r \sim -7.7$ \citep{Pota15} with $\sigma \sim 1.3$ mag \citep{Jordan07}, corresponding to an observed peak near $r \sim 20.5$. This means that the PISCeS survey, designed for depths of $r\sim26$--26.5, is quite effective at photometry of \emph{typical} GCs around NGC~5128. However, we have issues with recovering the most luminous GCs and ultra-compact dwarfs, which are up to $\sim10$ mag (a factor of 10000) brighter than our targeted individual red giants. PISCeS data start to saturate for point sources with $r \lesssim 18$ ($M_r \lesssim -10.2$) in the median field, with this limit pushed to $r \lesssim 18.6$--18.8 ($M_r \lesssim -9.6$ to --9.4) for the few pointings with the best seeing or highest background. For this reason, in our companion paper V20, we take a different approach to making a complete catalog of the most luminous sources, which relies only partially on the PISCeS data.

\subsection{{\it Gaia} Data Release 2}\label{sec:gaia_intro}

{\it Gaia} is an ongoing all-sky astrometric space mission. Its second data release (DR2) presents at least some astrometric and photometric measurements for over a billion sources down to a broadband $G \sim 21$ mag \citep{GaiaMain18}.

{\it Gaia} DR2 is a useful adjunct to the PISCeS data in two areas. First, through its measurements of proper motion and parallax, it allows us to identify and remove a large fraction of Galactic foreground stars that would otherwise contaminate our GC candidate sample. Second, with an effective angular resolution of $\sim 0.4\arcsec$
\citep{GaiaDR2validation}, {\it Gaia} DR2 includes astrometric and photometric statistics that can be used to pinpoint even marginally extended sources such as GCs in NGC~5128.

We explain our use of {\it Gaia} DR2 to identify NGC~5128 GCs in \S \ref{sec:gaia_dr2}.   The newly available {\it Gaia} Early Data Release 3 \citep{Gaia_edr3} and other future updates will provide additional and higher quality data for sources, increasing the accuracy with which we can identify nearby extragalactic GCs.

\subsection{NOAO Source Catalog}

The NOAO Source Catalog (NSC) is a unified collection of nearly all of the public data taken on the Dark Energy Camera (DECam) on the Blanco 4-m telescope at CTIO  
as well as the Mosaic3 imager on the 4m Mayall telescope at KPNO  
\citep{Nidever2018}.  It includes our area of interest around NGC~5128, with typical depths of $u \sim$ 23.3, $g \sim$ 21.7, $r \sim$ 21.2, $i \sim$ 21.2, and $z \sim 21.0$.
For this work we have used Data Release 1 of the NSC, although Data Release 2 has recently become available \citep{nsc_dr2}.

GCs have spectral energy distributions that are distinct from both individual foreground stars and from background galaxies with extended star formation histories. However, the single color available from PISCeS ($g-r$) does an inadequate job at separating these populations.  The multi-color photometry available in the NSC provides yet another method to identify and remove Milky Way foreground stars that would otherwise contaminate our GC candidate sample.  
We explain the use of the NSC to identify NGC~5128 GCs via their color in \S \ref{sec:color}.

To bring together data obtained under different protocols and by multiple PI's, \citet{Nidever2018} reprocessed the raw data with consistent quality control, selection, and calibration.  Astrometric calibration with $Gaia$ DR1 was done for each chip independently.  Photometric calibration was done with model magnitudes constructed for their filters from public all-sky catalogs and the \citet{Schlegel1998} reddening map for the extinction term.
There is a fairly consistent, color-dependent offset between the PISCeS and NSC photometry, between $\pm$ 0.1 mag in the $r$-band and 0.05 and 0.17 mag in the $g$-band.  We provide information about both PISCeS and NSC photometry in our final catalog of GC candidates.

The NSC includes the same data that comprises the Survey of Centaurus A's Baryonic Structures (SCABS), a wide-field $ugriz$
imaging survey of NGC~5128 using DECam 
\citep{Taylor2016,Taylor2017}, hereafter T17.    
T17 provides a list of likely NGC~5128 GC candidates out to $\sim$140 kpc based on color and some structural information.  We do not use the SCABS data catalogs directly because:  
(i) the PISCeS data generally have higher image quality (median FWHM = $0.66\arcsec$) than the SCABS data (FWHM $\sim 0.8\arcsec$--1.2\arcsec); 
(ii) it is clear from intercomparisons of PISCeS, NSC, and SCABS that the latter
suffers from field edge effects and significant, inconsistent magnitude offsets; and 
(iii) the NSC has a higher completeness for known GCs than SCABS.  We also note in passing that Table~2 of T17 appears to be extinction corrected, despite the statements in the table caption and text that it has not been (we hope this information helps future researchers using these data).
Within the PISCeS footprint, the NSC contains the reprocessed raw data from SCABS and other surveys with consistent quality control, selection, and calibration that results in a catalog with higher quality than SCABS alone.


\section{A fiducial sample of radial velocity-confirmed globular clusters} \label{sec:test sample}

To help guide and assess the fidelity of our GC candidate selection, we first build a fiducial sample of radial velocity-confirmed GCs in NGC~5128. We find an initial set of 630 ``confirmed" GCs by cross-matching published samples \citep{Graham1980, VanDenBergh1981, Harris1992,  Harris2002, Peng2004, Woodley2005, Harris2006,  Rejkuba2007, Woodley2007, Beasley2008, Georgiev2009, Woodley2010a, Woodley2010b, Taylor2010, Georgiev2010, Fahrion2020, Muller2020, Voggel2020}. The most common way to confirm a GC is based on its radial velocity, though some studies instead rely on morphological parameters or visual classification with \textit{Hubble Space Telescope} ({\it HST}) imaging. 
Updated {\it Gaia} DR2 and radial velocity measurements for many of these sources show that 73 objects are not actually GCs in NGC~5128; see Appendix \ref{app:misconfirmed} for details. Removing this contamination leaves a cleaned list of 557 confirmed GCs around NGC~5128, which we discuss and provide in Table \ref{table:confirmed_short} in Appendix \ref{app:confirmed}.

For the specific goals of this paper, we select a subset of these confirmed GCs to guide our selection process. 
Each GC in this fiducial sample:
(i) is $> 10\arcmin$ from the center of NGC~5128; 
(ii) has a radial velocity $> 250$ km s$^{-1}$; 
(iii) is in the {\it Gaia} DR2 catalog and does not have proper motion and parallax values consistent with being in the foreground;  
(iv) has not been marked as a foreground star or background galaxy in any catalog; and
(v) has photometry in our PISCeS {\tt Source Extractor} catalogs and in the NOAO Source Catalog.
This leaves a set of 69 fiducial GCs, as noted in Table \ref{table:confirmed_short} that can be used to test all of the criteria we use for selecting GCs outside the central regions of NGC~5128.  
We think nearly all of the GCs in the larger, cleaned sample of 557 confirmed objects are true GCs associated with NGC~5128, and use this broader list to assess the completeness of our sample in Section \ref{sec:compare}.

In addition to our fiducial sample of GCs, we also construct similar sets of foreground stars and background galaxies 
that have been confirmed via radial velocity measurements or \textit{HST} imaging, as found by previous studies of NGC~5128  \citep{Peng2004, Woodley2005, Harris2006, Gomez2006, Beasley2008}. These confirmed contaminant samples allow independent tests of the precision of our GC selection techniques.


\section{Identifying Globular Cluster Candidates}\label{sec:GCselection}

This section describes the process of finding GC candidates within the 95 calibrated images from the PISCeS survey, incorporating data from {\it Gaia} DR2 and NSC.  A summary is given in Section \ref{sec:summary}.        
. 

To assess the effectiveness of each step of our selection process, we compare our results with our fiducial sample of confirmed GCs and catalogs of known foreground stars and background galaxies.  
Information from each part of our selection process is used to calculate a final GC likelihood score between 0 and 1 for each candidate, where a score of 1 corresponds to those most likely to be true GCs in NGC~5128.  Where possible, we prioritize completeness over purity in each step of our selection process, though we recognize the limitations that our datasets have in regards to the brightest and faintest GCs.

\subsection{Initial Cuts}\label{sec:initial}

At the bright end, we remove saturated or nearly saturated sources, which typically have $r \lesssim 18$ mag (see also \S 4.2). 
We use a faint magnitude limit of 22 mag in the band ($g$ or $r$) with better seeing.  A magnitude of $g = 22$ mag is $\sim 0.7$ mag 
(a factor of $\sim 1.9$ in luminosity) beyond the turnover of the GCLF. Fainter than this limit, contamination grows substantially, and ancillary information from {\it Gaia} DR2 and NSC is either absent or of lower quality.  We use the same limit with the $r$ band ($r<22$ mag) to help remove spurious sources, since essentially all true GCs have $g-r > 0$. Finally, we restrict our study to projected distances greater than $10\arcmin$ from the center of NGC~5128, as crowding strongly affects our photometry in this inner region. Much of this central area is already covered by high-quality \emph{HST} data and other GC surveys.

We do not include in our candidate catalog any objects that have been confirmed in previous NGC~5128 studies as background galaxies or foreground stars via radial velocity measurements or \textit{HST} imaging, unless specifically noted in Table \ref{table:conflicting} in Appendix \ref{app:conflicting}.

\subsection{Concentration Index} \label{sub:2_ap_tech}

\begin{figure*}
\centering
\includegraphics[width=0.49\linewidth]{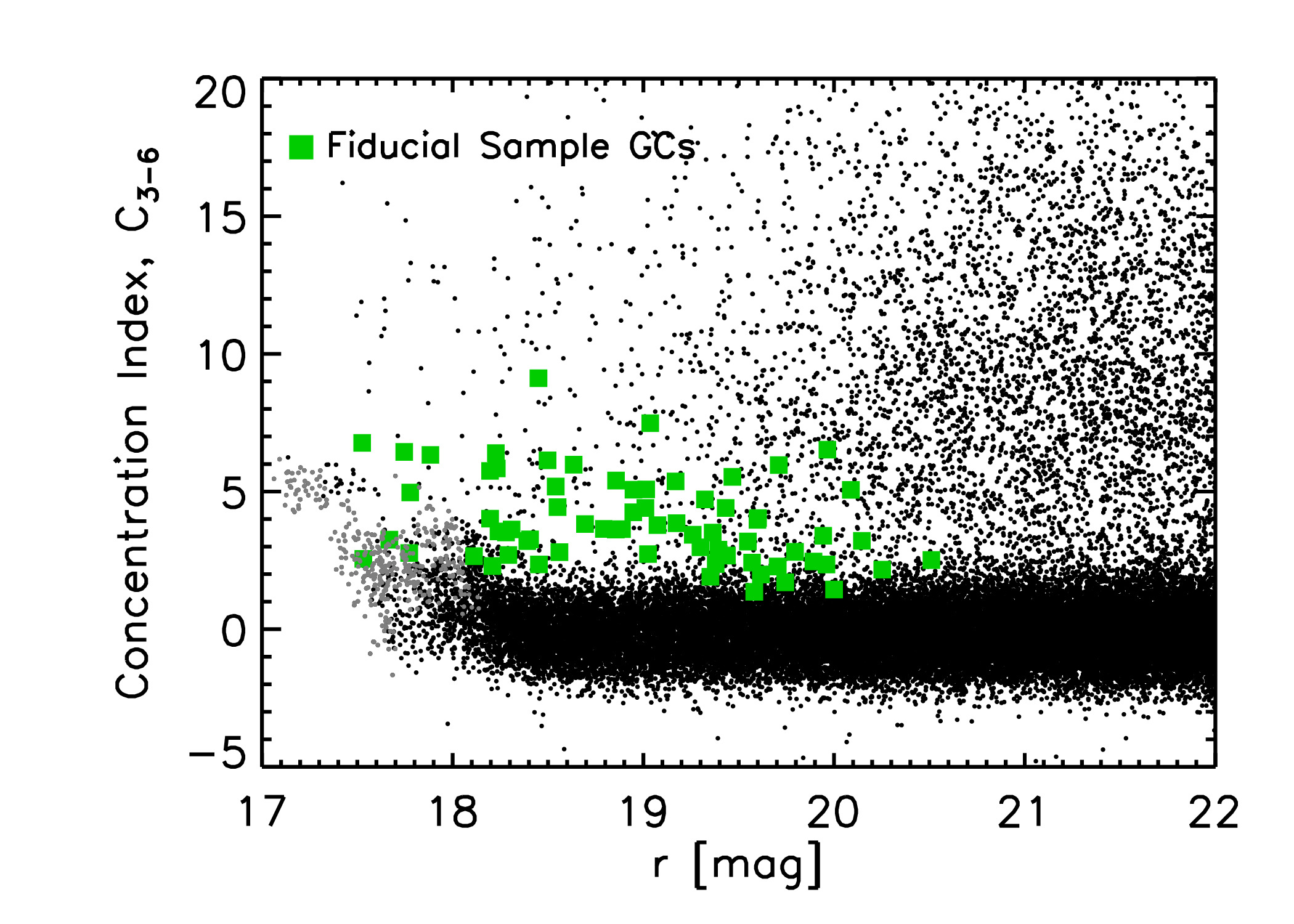}
~
\includegraphics[width=0.49\linewidth]{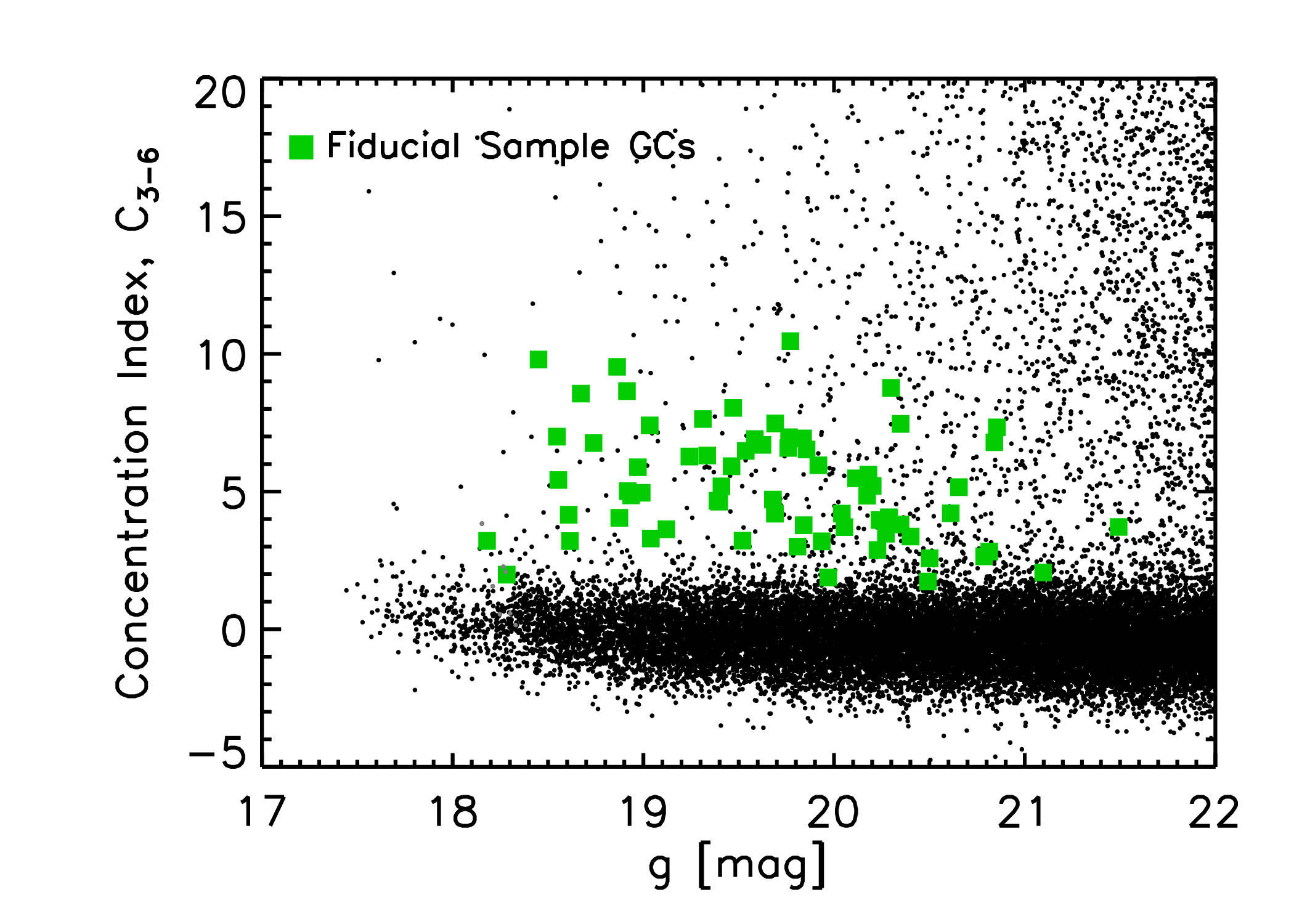}
\caption{ The concentration index ($C_{3-6}$) for fiducial sample GCs (green squares) and all sources between 10--30\arcmin~of the center of NGC~5128 (black dots). The left panel shows $r$-band data and the right panel shows $g$-band data. Objects with a larger $C_{3-6}$, such as GCs and background galaxies, are extended compared to point source foreground stars. Objects plotted in gray have been removed due to saturation/near saturation at brighter magnitudes.
}
\label{fig:magdiff}
\end{figure*}

Milky Way GCs range in half-light radius from 1--20 pc (with a mean of $\approx 3$ pc), corresponding to 0.05--1.08\arcsec~ at the distance of NGC~5128. In the excellent seeing of the 
PISCeS data (median FWHM of $0.66\arcsec$), this means that GCs are slightly extended compared to point sources. The most massive and largest GCs can be readily identified by the resolved stars along their outskirts (see also V20).

We can utilize the extended nature of GCs in NGC~5128 to separate them from foreground stars in the Milky Way using a concentration index--magnitude diagram.
This method has been commonly used in previous extragalactic GC studies to separate candidate GCs from foreground stars or background galaxies, both from the ground and with \emph{HST} \citep{Whitmore1999, Peng2011, Durrell2014, Peng2016, Beasley2016, Ko2019}. 

To calculate the concentration indices, for each field we measure the difference in magnitudes at diameters of 3 and 6 pixels (0.48\arcsec\ and 0.96\arcsec, respectively) for each source. These aperture sizes were chosen to be effective at separating true GCs from contaminants under a range of crowding and signal-to-noise conditions. Because the image quality differs among the 95 fields, the mean magnitude difference for foreground stars varies between 0.72 and 1.28 mag, with a standard deviation in the range 0.01--0.07 mag. To create a concentration index that can be compared among fields, we select unsaturated point sources with magnitudes between 18.5 and 20.0, and normalize the distribution of magnitude differences to a mean of 0 and a standard deviation of 1 on a per field basis after trimming outliers.\footnote{Using the IDL code \texttt{resistant$\_$mean}. All robust means and standard deviations calculated in this paper use the same code.}
Figure \ref{fig:magdiff} shows the concentration index--magnitude diagram for all sources between 10 and 30\arcmin~of NGC~5128, highlighting the separation of the fiducial sample GCs from foreground stars. We use the symbol $C_{3-6}$ to represent these normalized values, where larger $C_{3-6}$ values correspond to more extended sources.

Assuming that point sources are distributed in a Gaussian distribution centered around $C_{3-6} = 0$,  \textit{concentration\_likelihood} for each GC candidate is defined  
as 1 minus the probability that the candidate's $C_{3-6}$ value is consistent with a point source.  For this calculation, we use data from the observing band ($g$ or $r$) with better seeing.
As an example, if a candidate has $C_{3-6} = 2.0$, it lies 2$\sigma$ above the point source average for that field, and we assign it a \textit{concentration\_likelihood} of $0.977$.

Stars that are nearly saturated lead to $C_{3-6}$ values above zero at bright magnitudes even for true point sources (see stars with $r \lesssim 18$ in Figure~\ref{fig:magdiff}). We exclude these saturated (or nearly saturated) objects. Reliably identifying the most luminous GCs and ultra-compact dwarfs requires a separate set of selection criteria, discussed in V20.

All of the fiducial sample GCs have $C_{3-6}$ values greater than 1.3 in the $r$-band and greater than 1.7 in the $g$-band, corresponding to \textit{concentration\_likelihood} values greater than 0.912 and 0.958, respectively.  
The average $C_{3-6}$ value of the fiducial GCs is 3.9 in the $r$-band, and 5.2 in the $g$-band.
For each candidate we use the band with better seeing to calculate its likelihood, noting that the fiducial sample is recovered with high purity in both bands.

We compiled all available size information for GCs in NGC~5128 \citep{McLaughlin2008, Rejkuba2007, Taylor2010, Taylor2015}
Only 16 of the 69 fiducial sample GCs have physical size information, with half-light radii between $\approx$~2.1--4.9 pc and an average of 3.0 pc.  The smallest of these has a concentration index of 1.7 in the $r$-band and 4.1 in the $g$-band.  Because most searches are not sensitive to the most compact GCs and
our selection cuts are based on previously known GCs, our final catalog may also not include the smallest GCs (half-light radius $\approx$~1--2 pc) in NGC~5128.

All sources in the 95 PISCeS fields that passed the initial cuts and have a \textit{concentration\_likelihood} $>$ 0.84 ($C_{3-6} > 1.0$) are passed as GC candidates to the next stage of the selection process.


\subsection {Removing Galaxies and Blends} \label{sec:galcul}  

\begin{figure*}[ht]
\hspace{0.4cm}
\begin{minipage}[h]{0.4\linewidth}
\centering
    \includegraphics[width=\linewidth]{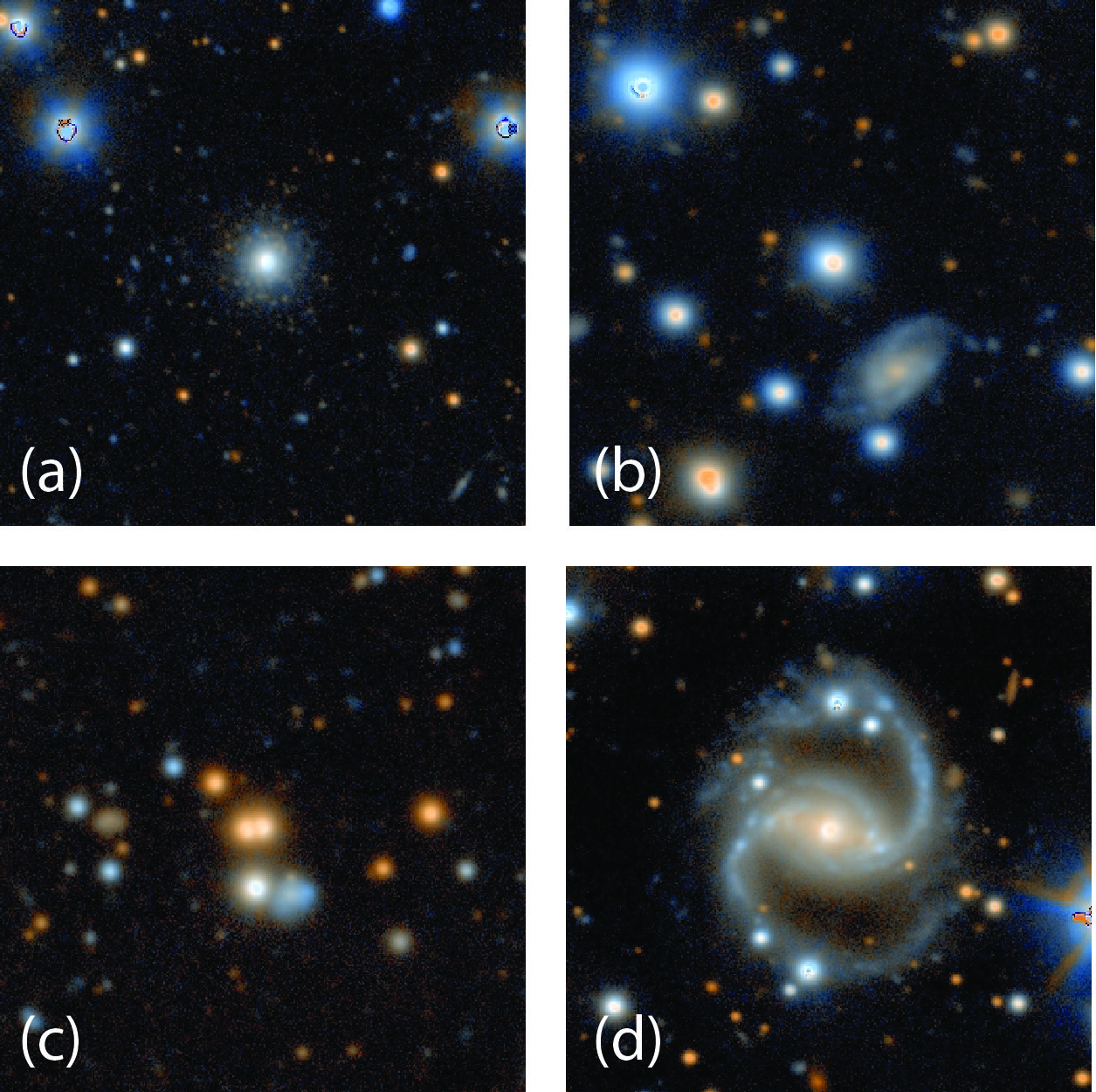}
\end{minipage}
\hspace{0.4cm}
\begin{minipage}[h]{0.51\linewidth}
\centering
\includegraphics[width=\textwidth]{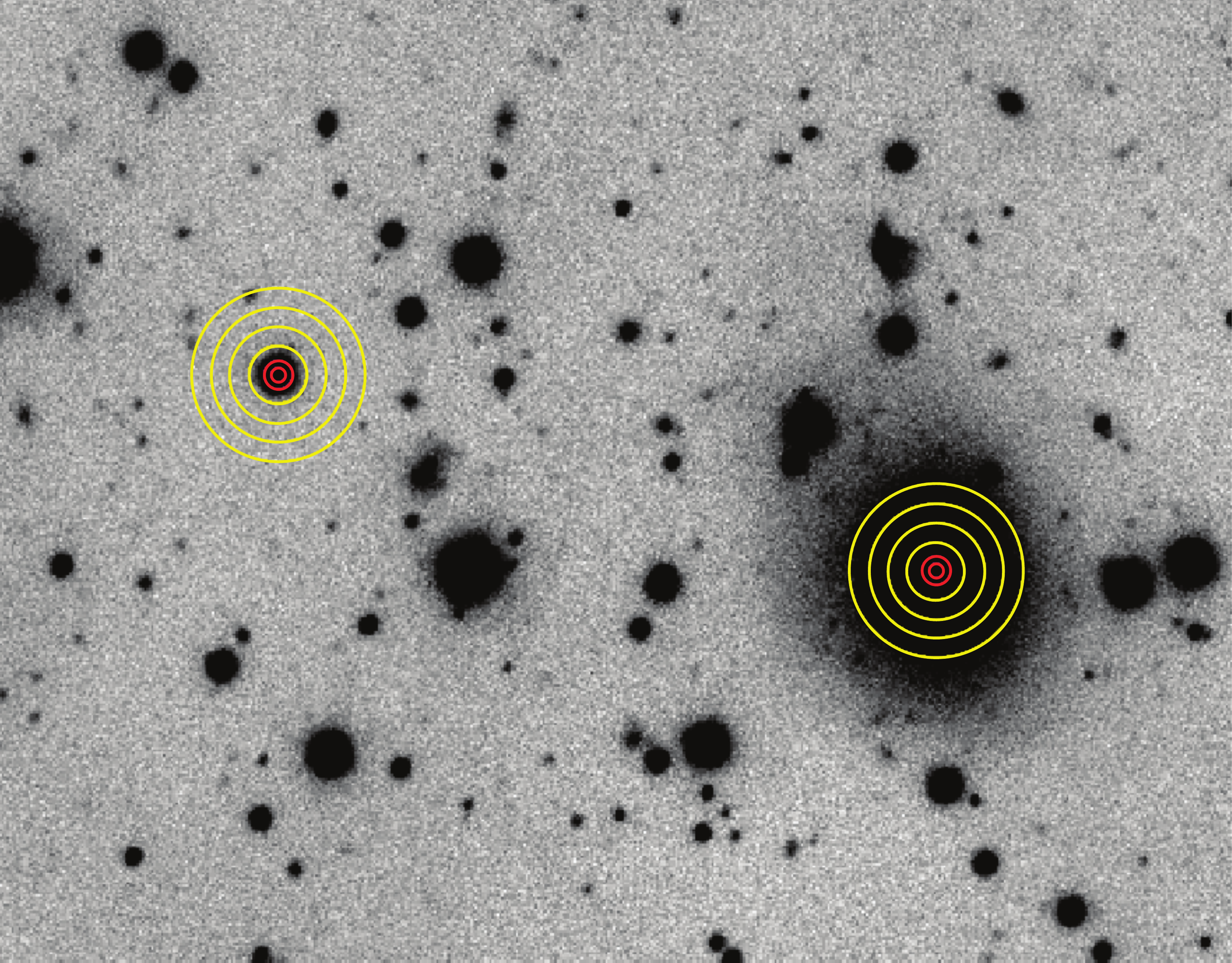}
\end{minipage}
\caption{Examples of extended objects with  $C_{3-6} >$ 3.  In the left panel, starting in the upper left subpanel and going clockwise is: (a) A confirmed, luminous GC (KV19-442) that is partially resolved, with red giants apparent in its outskirts; (b) a good GC candidate that is not resolved into stars, but which is extended compared to a point source; (c) a pair of foreground stars that are close enough that they are photometered as a single source with a large $C_{3-6}$ value; and (d) a very extended galaxy that is cut during the galaxy removal process (\S 4.3).  Each image is $\approx 1' \times 1'$.
In the right panel, we show an example of how a GC can be distinguished from a background galaxy by using a brightness profile compiled from a series of concentric apertures with 12, 20, 28, and 36 pixel diameters (yellow circles). The smaller, red circles show the 3 and 6 pixel diameter apertures used for the concentration index measurements.}
\label{fig:visual_examples}
\end{figure*}

Besides GCs, there are other types of extended sources in the PISCeS data.
Among objects with higher values of concentration index ($C_{3-6} > 3$), there is a substantial contribution from background galaxies, especially toward fainter magnitudes. Due to the relatively low Galactic latitude of NGC~5128 ($b= 19^{\circ}$) blends or near-blends of foreground stars are also common. Figure \ref{fig:visual_examples} shows visual examples of both true GCs and common contaminants.

The process described in this section aims to reduce contamination from background galaxies and stellar blends in our final GC candidate catalog  by analyzing the shape and brightness profile of each of our potential GC candidate targets. 

We build an algorithm that produces a binary output, identifying a source as being either: (1) too elliptical and/or too extended to be a GC in NGC~5128, in which case it is given a structure-based likelihood (\textit{structure\_check}) of 0, or (2) round and only slightly extended, as expected for GCs, in which case it is given \textit{structure\_check} = 1.  This algorithm was designed empirically such that a majority of visually identifiable galaxies receive \textit{structure\_check} = 0, while the fiducial sample GCs receive \textit{structure\_check} = 1.  

The galaxy/blend assessment that produces this final binary output is a three-step process. For each source, we first take the Source Extractor \texttt{flux\_radius} parameter, which measures the circular-equivalent effective radius (the circular aperture radius enclosing half the total flux of an object) in units of pixels. We then compare this to a conservative limit, which is a robust average \texttt{flux\_radius} of foreground stars in the field + 3 pixels, tuned to pass even marginal GC candidates and exclude the more obvious galaxies. Objects above this critical value are given \textit{structure\_check} = 0, while those with \texttt{flux\_radius} values below the limit continue to the next step.

We next check the ellipticity of the candidates, motivated by the fact that true GCs are mostly round, while background galaxies (and blends) have a broader ellipticity distribution. The ellipticity $\epsilon$ is defined as $\epsilon = 1-b/ a$, where $a$ and $b$ are the semi-major and semi-minor axes measured by {\tt Source Extractor}. We reject the source (set \textit{structure\_check} = 0) if $\epsilon > 0.35$ in one of $g$- and $r$-bands and $\epsilon > 0.25$ in the other. 
A few of the most massive star clusters around NGC~5128 have ellipticities up to $\epsilon \sim 0.35$, likely related to their origin as stripped galaxy nuclei \citep{Voggel18}. Hence our ellipticity cut could in principle remove a small number of sources truly associated with NGC~5128. 
Since the current paper is focused on typical GCs rather than the most massive objects, this ellipticity cut represents an important step in removing the contaminants that are abundant at fainter magnitudes.

As a final evaluation of how extended the brightness profiles of individual candidates are, in each band we measure the magnitude of each target in concentric apertures of diameter 12, 20, 28, and 36 pixels, and determine the magnitude differences between successive pairs (diff\_12\_20, diff\_20\_28, and
diff\_28\_36). For this step, we reject the source (set \textit{structure\_check} = 0) 
as too extended if \emph{all} three of the following conditions are met in either $g$- or $r$-band: (i) diff\_12\_20 is 10 times greater than the aperture correction value (see Section \ref{sub:pisces}); (ii) 
diff\_20\_28 $>$ 0.1; and (iii) diff\_28\_36 $>$ 0.1. These larger apertures effectively remove extended, moderately round background galaxies that passed the previous selection steps. See Figure \ref{fig:visual_examples} for a visual example of this process.

As a test of this galaxy/blend removal step, we compare our results with the velocity-confirmed background galaxies from several previous spectroscopic studies of GCs around NGC~5128 \citep{Peng2004, Woodley2005, Gomez2006, Beasley2008}. These sources were originally selected in a variety of ways, but generally had colors and magnitudes consistent with those expected for GCs.
We find that the steps in this subsection
identify 58 out of 95 (61\%) unique background galaxies from these previous studies. 
It is reasonable to assume that a subset of the more obvious background galaxies were culled prior to spectroscopy in these studies. Therefore, our selection steps can be expected to successfully flag and remove a large fraction of background galaxies in our magnitude range.

By examining the confirmed background galaxies \emph{not} identified by our selection steps, we find that these are
marginally extended and circular and hence
visually similar to true GCs. As we discuss below, additional photometric or spectroscopic information is
needed to reliably reject these remaining background galaxies.

In total, 66\% of the possible GC candidates that meet the criteria detailed at the end of \S 4.2 are given \textit{structure\_check} = 1, and the other 34\% are given \textit{structure\_check} = 0 and removed from the GC candidate sample.


\subsection{{\it Gaia} DR2} \label{sec:gaia_dr2}

Section~\ref{sec:gaia_intro} briefly introduced the possibility of using 
{\it Gaia} DR2 to improve our GC catalog both in relatively straightforward ways (by removing foreground stars) and in a more unexpected fashion (by identifying marginally resolved sources). Here we discuss how we use a range of astrometric and photometric {\it Gaia} DR2 measurements in our GC candidate selection process.

First, we note that 305 of 
the 557 confirmed NGC~5128 GCs (55\%) are in {\it Gaia} DR2. About 80\% of the missing objects are faint ($g > 20$ in PISCeS or NSC) and likely below the DR2 completeness limit \citep{Boubert20}. Most of the rest are within $10\arcmin$ of the galaxy center (where photometry is difficult), or have conflicting literature classifications (see Appendix \ref{app:conflicting} for details).

Of the GC candidates that passed the galaxy/blend removal step in \S 4.3, we find that 62\% of these sources are in {\it Gaia} DR2. 
If a source is \emph{not} in DR2, it is simply passed to the color selection in \S 4.5.  Likewise, if a source is in {\it Gaia} DR2, but has a missing measurement (such as the BP/RP excess factor discussed below), we simply skip the evaluation of the relevant step.

\subsubsection{Proper Motion \& Parallax}

The proper motions and parallaxes of objects associated with NGC~5128 cannot be measured with {\it Gaia} DR2. Therefore, any sources with significant proper motions or parallaxes can be identified as being in the foreground, and removed from the GC candidate sample.

We first calculate the total proper motion of an object by adding the right ascension and declination components in quadrature. Assuming the combined proper motion has a Chi distribution with two degrees of freedom, it can be simplified to a Rayleigh distribution. 
We assign a \textit{proper\_motion\_likelihood} of 1--CDF, where CDF is the cumulative distribution function of the Rayleigh distribution.  Sources with lower \textit{proper\_motion\_likelihood} values are therefore more likely to be foreground stars based on their proper motion measurements.
For example, if the proper motion is significant at $2\sigma$, the likelihood assigned is 1--0.865 = 0.135. We truncate this at $3\sigma$ (0.011): sources with proper motions more significant than this value are given \textit{proper\_motion\_likelihood} = 0 and removed from the candidate list.

We also calculate a similar \textit{parallax\_likelihood}  based on the significance of the parallax, here assuming the measurement has a Gaussian distribution. For example, if the parallax is significant at $2\sigma$, the likelihood assigned is $1 - 0.954 = 0.046$. Sources with $> 3\sigma$ parallax measurements are given  \textit{parallax\_likelihood} = 0 and removed from the candidate list. Since significant parallax measurements are much less common than significant proper motion measurements in {\it Gaia} DR2, it is rare to have an object with a significant parallax but not a significant proper motion.

Of the GC candidates in {\it Gaia} DR2 that passed the galaxy/blend removal step in \S 4.3, we find that 68\% are removed for having proper motion and/or parallax measurements consistent with being foreground stars.   

Of the 391 confirmed foreground stars in previously published papers, 313 (80\%) are identified using these \textit{proper\_motion\_likelihood} and \textit{parallax\_likelihood} criteria. The remainder have insignificant proper motions or parallaxes or are not in {\it Gaia} DR2, primarily because they are faint. This comparison shows that {\it Gaia} is an effective tool to identify and remove foreground stars from our GC candidate sample.


\subsubsection{Astrometric Excess Noise}

\begin{figure*}
\centering
\includegraphics[width=0.48\linewidth]{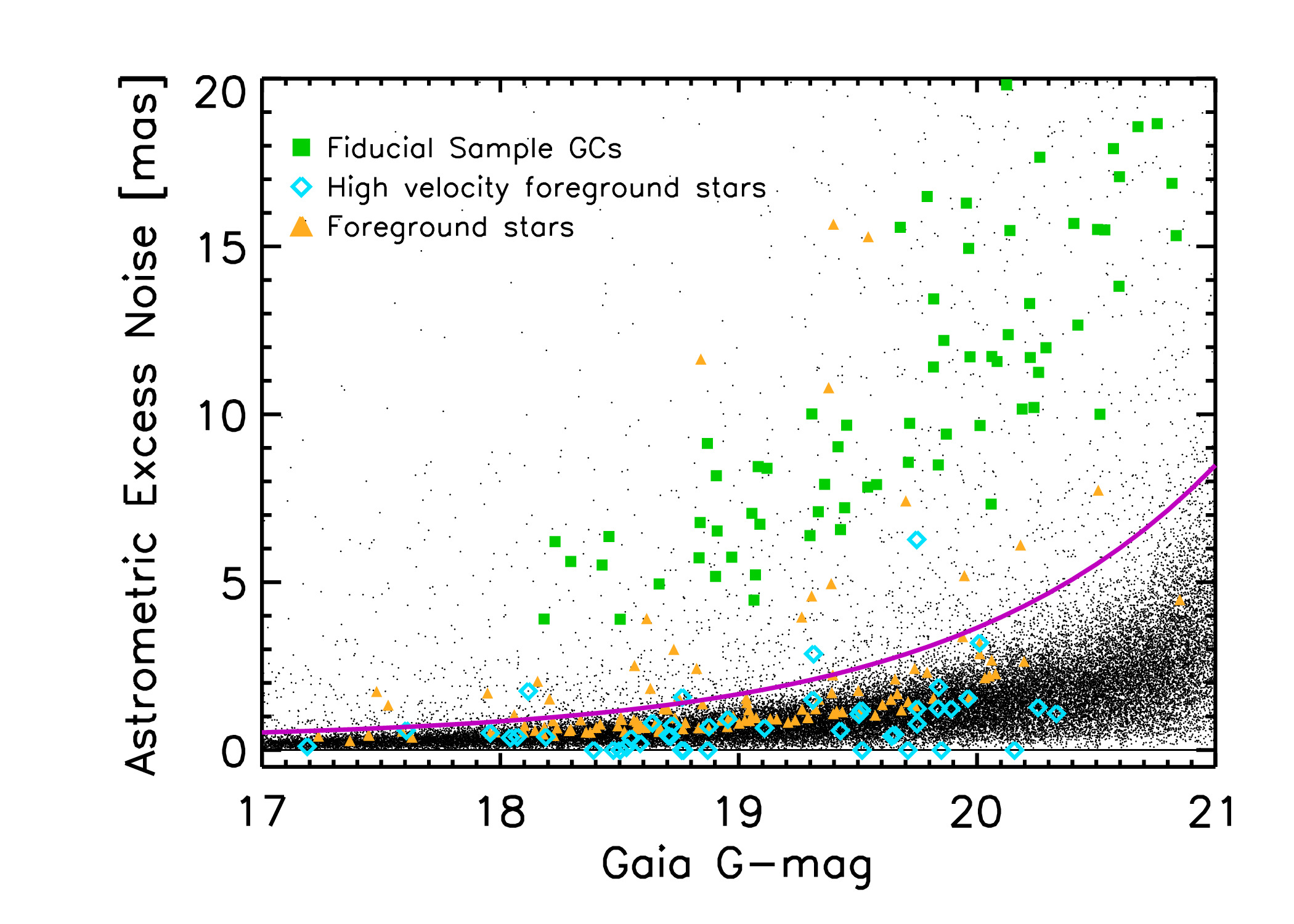}
~
\includegraphics[width=0.48\linewidth]{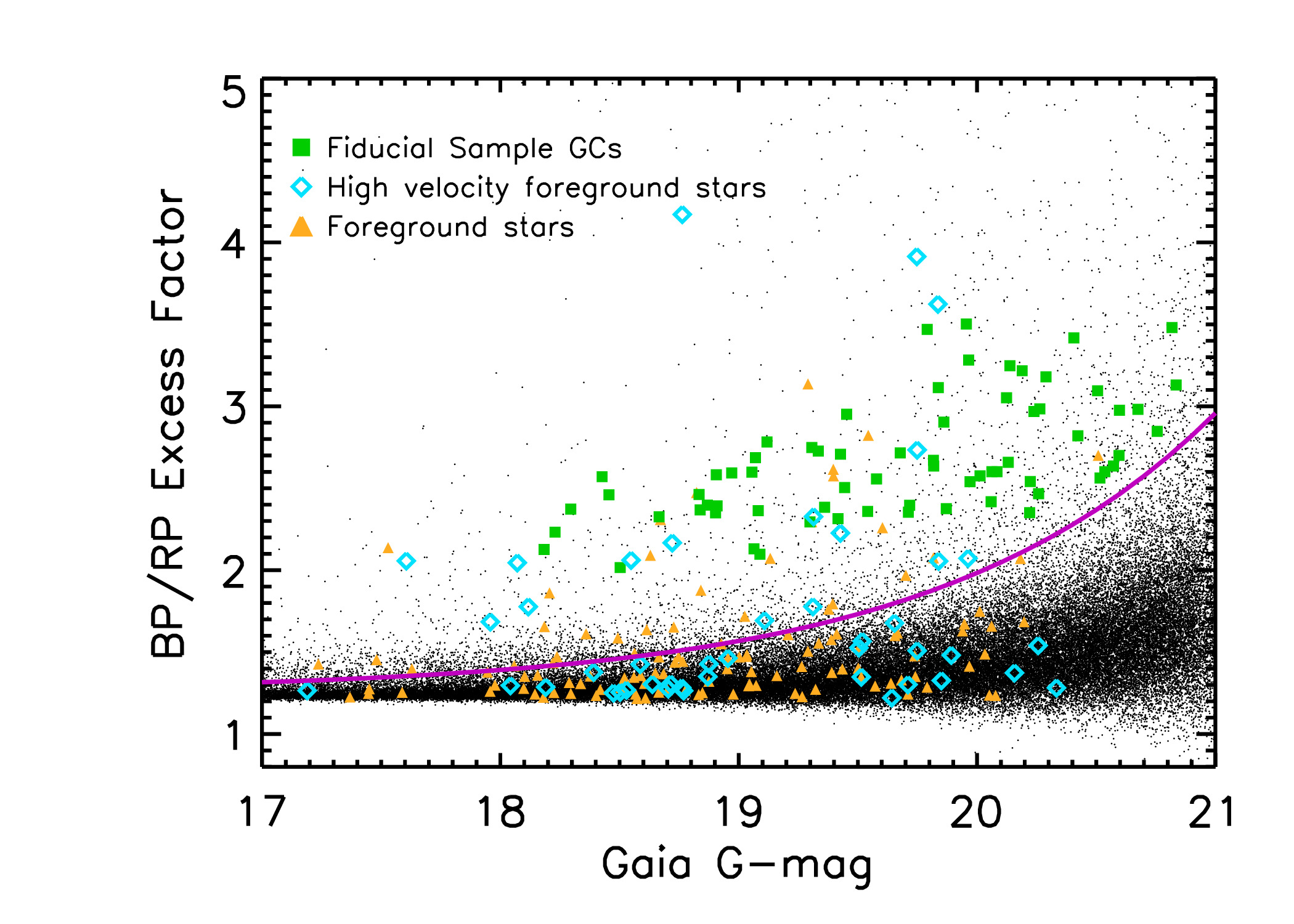}

\caption{Left: Plot of Gaia's AEN parameter versus G band magnitude.   Right: Plot of \textit{Gaia}'s $BR_{\rm excess}$ parameter versus G band magnitude.  In both plots, the black dots are all {\it Gaia} DR2 sources between 10--60\arcmin~of NGC~5128, the green squares are fiducial sample GCs, blue diamonds are foreground stars based on their proper motion (discussed in Appendix \ref{app:hvfs}), and orange triangles are radial velocity confirmed foreground stars.
The fiducial sample GCs are distinct from the general population of foreground stars in both parameter-spaces, and so we can use the AEN and $BR_{\rm excess}$ parameters to identify GC candidates. The lines denoting 3$\sigma$ are given by Equations 1 and 2. Confirmed foreground stars with AEN or $BR_{\rm excess}$ values greater than 3$\sigma$ are stellar blends, visible in PISCeS images.}
\label{fig:aen_bprp}
\end{figure*}

{\it Gaia} DR2 reports the astrometric excess noise (AEN) as a statistic that quantifies the goodness-of-fit of their five-parameter astrometric model to the astrometry for each target \citep{Gaia_summary}. The AEN, given in units of milli-arcseconds (mas), is equal to 0 for a well-fit star and has larger values for poorer fits. Objects that are extended compared to point sources have large AEN values, and in V20 we showed that this can be used to select marginally resolved objects such as GCs around NGC~5128.  

In Figure \ref{fig:aen_bprp} we plot AEN-{\it Gaia} DR2 $G$ mag for sources around NGC~5128. Most of these are
foreground stars and therefore have low AEN values ($\lesssim 1$), especially at brighter magnitudes where the random uncertainties are smaller. This figure also plots the fiducial sample GCs, showing that they have larger AEN values and are largely distinct from foreground stars. Much like the concentration index $C_{3-6}$, AEN is more valuable for rejecting foreground stars than for rejecting extended objects such as background galaxies or foreground star blends, which like GCs will have higher AEN values.

Because the AEN distribution depends on the {\it Gaia} DR2 $G$ mag, we determined robust AEN mean and standard deviation for point sources in 0.1 mag bins. We then calculate a candidate's \textit{aen\_likelihood} as 1 minus the probability that its AEN value is consistent with a point source at that magnitude, truncating the distribution at $-2\sigma$ (\textit{aen\_likelihood} = 0.023). 
Rather than an error, each AEN value is also reported with a corresponding ``significance" measure, where a  value greater than 2 indicates that the given AEN is probably significant. Because a majority of sources with high AEN values are not single stars, they have bad astrometric fits and therefore high significance values.

As an example calculation, consider a GC candidate with an AEN value of 2.14 mas and {\it Gaia} DR2 $G = 20.03$ mag. In the bin between $G=20.0$ and 20.1 mag, foreground stars have an AEN robust mean of 1.465 mas and $\sigma = 0.672$ mas. This GC candidate is $1\sigma$ above the mean for point sources and therefore has an \textit{aen\_likelihood} of 0.841. 
 
For the convenience of the reader who wishes to have a simple criterion to find extended sources in other circumstances, we use information from {\it Gaia} DR2 for sources in the radial range 10--60\arcmin\ from the galaxy center to define a line 3$\sigma$ above the mean foreground star AEN value: 
\begin{equation}
AEN_{3\sigma} = 0.297 + 5.63\times 10^{-8} \, e^{0.895\,G}
\label{eqn:aen_3sig}
\end{equation}
where $G$ is the {\it Gaia} DR2 $G$ mag. This line is plotted in Figure \ref{fig:aen_bprp}. We emphasize that
for this paper we use the likelihood analysis described above.

All of the 69 fiducial sample GCs are greater than 4$\sigma$ above the mean AEN for foreground stars at their magnitude. Of the 375 confirmed foreground stars in {\it Gaia} DR2, 345 (92\%) have AEN values within 3$\sigma$ of the mean for their $G$ magnitude. Visual inspection using PISCeS imaging shows that the remaining 30 stars with large AEN values are all close blends. This confirms the utility of AEN for rejecting single foreground stars from the GC candidate catalog.


\subsubsection{BP/RP Excess Factor}

As we previously showed in V20, a second {\it Gaia} DR2 parameter useful for finding extended sources is the BP/RP Excess Factor ($BR_{\rm excess}$).
This parameter gives the ratio of the sum of the flux of the Blue Photometer ($BP$, 3300-6800 \AA) plus the Red Photometer ($RP$, 6400-10500 \AA) with the flux in the broadband $G$ filter. Its utility comes from the distinct manner in which these fluxes are determined: the $BP$ and $RP$ magnitudes are measured directly from the flux within a large aperture of 3.4 $\times$ 2.1 arcsec$^2$, while the $G$ magnitude is derived from profile fitting with an effective resolution of $\sim$0.4\arcsec\ \citep{Evans2018}. This difference means that extended sources have a higher $BR_{\rm excess}$ than point sources, since fitting an effective point-source profile to an extended source in $G$ misses some of the light.
As for the concentration index $C_{3-6}$ and AEN, we expect all extended objects, including contaminant background galaxies and foreground star blends, to have large $BR_{\rm excess}$ values, while Galactic foreground stars will have small values. 

Figure \ref{fig:aen_bprp} (right panel) shows $BR_{\rm excess}$ plotted against {\it Gaia} DR2 $G$ mag. Qualitatively the figure is similar to the corresponding one for AEN, with elevated  
$BR_{\rm excess}$ values for the fiducial sample of GCs and a clear point-source track of foreground stars. The 
fiducial GCs are less clearly separated from the stars than for AEN, but the data used are essentially independent and so it serves as a second valuable piece of information to find extended sources. We calculate a \textit{bre\_likelihood} for each candidate in the same manner as for \textit{aen\_likelihood}, described in \S 4.4.2.

We also constructed a similar line 3$\sigma$ 
above the mean foreground star $BR_{\rm excess}$ value:
\begin{equation}
BR_{excess, 3\sigma} = 1.26 + 2.79\times 10^{-8} \, e^{0.853\,G}
\label{eqn:bprp_3sig}
\end{equation}
where G is the {\it Gaia} DR2 $G$ mag. 

All of the fiducial sample GCs have a $BR_{\rm excess}$ value at least 3$\sigma$ above the mean $BR_{\rm excess}$ for foreground stars at their magnitude, and 90\% of them have a $BR_{\rm excess}$ value above 4$\sigma$, excluding a single fiducial GC that does not have a $BR_{\rm excess}$ measurement.

Of the 367 confirmed foreground stars with $BR_{\rm excess}$ data, 300 (80\%) have $BR_{\rm excess}$ values within 3$\sigma$ of the mean for their magnitude.
Reflecting what is seen with AEN, the remainder are mostly foreground star blends.

While both AEN and $BR_{\rm excess}$ measure how extended sources are in {\it Gaia} DR2---and hence are correlated for truly extended sources---the statistics are largely independent and hence both valuable.


\subsection{Color Selection}\label{sec:color}

As old stellar populations with typically simple star formation histories, GCs have a restricted range of spectral energy distributions, governed primarily by metallicity. A fortunate side effect is that GCs occupy a limited area of parameter space in optical color--color diagrams, allowing them to be separated from foreground stars and background galaxies with reasonable fidelity.
Hence color--color diagrams have a long history in 
the selection of extragalactic GCs (e.g., \citealt{Rhode01,Strader2011,Jennings14,Brodie14,Powalka16a}; T17). The use of a wide spectral baseline, such as including the UV, near-IR, or both, improves the purity of the GC selection \citep{Munoz2014,Powalka16b}.

Since our PISCeS data have only two filters ($g$ and $r$ band), to use the color selection method we match our catalog with NSC, which has \textit{ugriz} photometry for over 100,000 sources within the PISCeS footprint \citep{Nidever2018}.   
Following the methods of past GC selection techniques, we use the $u$, $r$, and $z$ filters to maximise the separation of GCs from foreground stars.

\begin{figure}[t]
\includegraphics[width=1.0\linewidth]{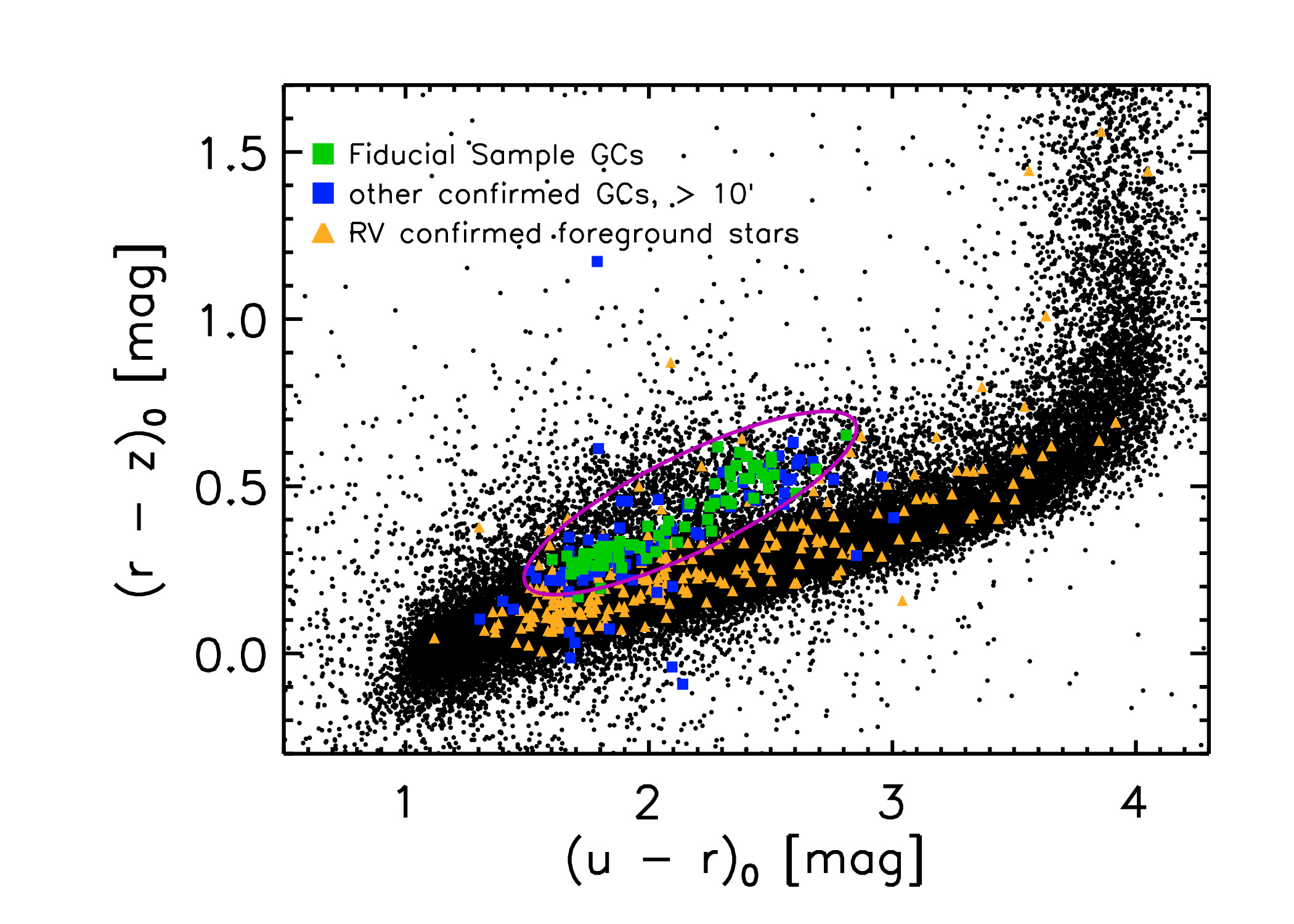}
\caption{Optical color-color diagram of all sources at distances between 10 and 60\arcmin of NGC~5128 that are in both PISCeS and NSC.  Almost all of the fiducial sample GCs (green squares) lie in a well-defined region that is separate from the radial velocity confirmed foreground stars (orange triangles). The purple ellipse marks the elliptical region used to select GC candidates, which was chosen to bound the fiducial sample GCs and confirmed GCs at a distance greater than 10\arcmin~from the center of the galaxy. 
}
\label{fig:cct}
\end{figure}

Figure~\ref{fig:cct} shows the position in ($u-r$)$_0$--($r-z$)$_0$ space of all sources in the radial range 10\arcmin--60\arcmin ~from the center of NGC~5128 that are in both PISCeS and NSC.  The subscript 0 (e.g., ~$u_0$) denotes that the photometry has been corrected for foreground galactic extinction.
For the purposes of calculating a likelihood, we define the high-probability GC area as an ellipse 
centered at $(u-r)_0 = 2.17$ and  $(r-z)_0=0.45$, with semi-major axis 0.72 mag, semi-minor axis 0.15 mag, and a position angle $19^{\circ}$ from the x-axis, as illustrated in Figure~\ref{fig:cct}.
This ellipse encloses most of the fiducial GCs and the confirmed GCs outside of 10\arcmin~ from the center of the galaxy.  
Using integrated single stellar population magnitudes from \citet{Marigo2008}, we find that the cutoff in color on the blue side of the ellipse corresponds to GC models with ages $\gtrsim$ 5 Gyr at a [Fe/H] = --2, and $\gtrsim$ 1--2 Gyr for more metal rich stellar populations.

To calculate the color-based likelihood, we draw 10,000 random samples from the $u$, $r$, and $z$ bands for each source, assuming Gaussian photometric uncertainties. The fraction of these samples that lie within the ellipse is the \textit{color\_likelihood}. If all samples are within the ellipse, $\textit{color\_likelihood} = 1$. Since a few fiducial objects sit just outside the ellipse, and a few true GCs could have abnormal colors (e.g., due to a problem with their photometry in a single band), we set the minimum likelihood at $2\sigma$ (0.023), as we did for some of the Gaia~DR2-based likelihoods. Candidates not in the NSC do not have their total likelihood values affected by this step.

As an example calculation, let us look at a typical GC candidate with $u_0 = 21.42 \pm 0.12$, $r_0 = 18.81 \pm 0.01$, and $z_0 = 18.28 \pm 0.03$. This candidate's position in color-color space of $(u-r)_0 = 2.61$ and $(r-z)_0 = 0.53$ is inside of the color selection ellipse, but only 80\% of the randomly varied samples drawn are within the ellipse.  Therefore, it has \textit{color\_likelihood} = 0.80.

Out of the 69 fiducial sample GCs, 64 (93\%) have \textit{color\_likelihood} $>$ 0.95. Of all the potential GC candidates that pass the concentration index, galaxy, and proper motion/parallax tests, 29\% have NSC photometry, and of this 16\% have \textit{color\_likelihood} $>$ 0.75 and 4\% have \textit{color\_likelihood} = 1.0. This suggests the NSC color information is very effective in reducing remaining contaminants in the GC candidate catalog.


\subsection{Summary of GC candidate selection process}\label{sec:summary}

To summarize, there are five steps in our GC candidate selection process for the 95 PISCeS fields. We first exclude the brightest and faintest candidates and those affected by crowding near the galaxy's center. We then do a likelihood-based selection of extended objects using a two-aperture technique on the PISCeS photometry. In the next step, background galaxies and stellar blends are rejected using cuts in the effective radius, ellipticity, and the flux distribution at large radii.
The penultimate step uses {\it Gaia} DR2 for a likelihood-based assessment that a source is a GC rather than a foreground star based on measurements of its astrometric motion (proper motion and parallax) and whether it is extended (astrometric excess noise and BP/RP excess). Finally, we use multi-band photometry from the NSC to assign a likelihood that each source has colors consistent with confirmed GCs. 

The above likelihoods are multiplied together to give the \textit{total\_likelihood} of each candidate. Numbers closer to 1 represent a larger chance that the source is a GC, while candidates that did not survive a yes/no cut (i.e., those assigned a likelihood of 0 at any point) have a \textit{total\_likelihood} = 0. These latter objects do not appear in our final GC catalog.

The different steps do not have their likelihoods normalized, and no penalties are applied if a target does not appear in a given catalog. For instance, if a GC candidate is not in the {\it Gaia} DR2 catalog because it is too faint, it is implicitly given a likelihood of 1 at that step. The interpretation of sources with different supporting data is considered in the next section, where these likelihoods are discussed in the context of our full PISCeS catalog.


\section{Catalogs}\label{sec:catsec}
In this section, we present two catalogs: the first includes all of our GC candidates and the second includes all PISCeS sources that pass the initial cuts (see Section~\ref{sec:initial}).  We discuss these catalogs and how we rank our candidates based on the amount of data available for each.  

\subsection{GC Candidate and Overall Source Catalogs} \label{sec:catalog}

\begin{table*}[ht]
\centering
\caption{Header information for the catalog of GC candidates in NGC~5128}
\begin{tabular}{c c c}
\hline \hline

Column & Label & Description \\
\hline
1 & ID & unique `H21' ID  \\
2 & Discovery\_ID & Discovery ID, if available  \\
3 & RA & PISCeS R.A. in decimal degrees (J2000)  \\
4 & Dec & PISCeS Dec in decimal degrees (J2000)  \\
5 & PISCeS\_gmag & PISCeS $g$ band magnitude (mag) \\
6 & PISCeS\_gmag\_err & error of PISCeS $g$ band magnitude (mag)  \\
7 & PISCeS\_rmag & PISCeS $r$ band magnitude (mag) \\
8 & PISCeS\_rmag\_err & error of PISCeS $r$ band magnitude (mag)  \\
9 & C\_3\_6 & Concentration Index value  \\
10 & Gaia\_Gmag & Gaia DR2 $G$ band magnitude (mag) \\
11 & pm\_ra & Gaia DR2 proper motion measurement in R.A. direction (mas/yr) \\
12 & pm\_ra\_err & error of Gaia DR2 proper motion measurement in R.A. direction (mas/yr) \\
13 & pm\_dec & Gaia DR2 proper motion measurement in Dec direction (mas/yr) \\
14 & pm\_dec\_err & error of Gaia DR2 proper motion measurement in Dec direction (mas/yr) \\
15 & parallax & Gaia DR2 parallax measurement (mas) \\
16 & parallax\_err & error in Gaia DR2 parallax measurement (mas) \\
17 & AEN & Gaia DR2 Astrometric Excess Noise (mas) \\
18 & BR\_excess & Gaia DR2 BP/RP Excess Factor  \\
19 & ur\_color & NSC $(u - r)_0$, with Milky Way dust correction applied (mag)  \\
20 & rz\_color & NSC $(r - z)_0$, with Milky Way dust correction applied (mag) \\
21 & concentration\_likelihood & candidate likelihood based on $C_{3-6}$ \\
22 & proper\_motion\_likelihood & candidate likelihood based on proper motion \\
23 & parallax\_likelihood & candidate likelihood based on parallax \\
24 & aen\_likelihood & candidate likelihood basd on AEN \\
25 & bre\_likelihood & candidate likelihood based on BR$_{\rm excess}$ \\
26 & color\_likelihood & candidate likelihood based on NSC color \\
27 & total\_likelihood & total likelihood of candidate being a GC in NGC~5128 \\
28 & candidate\_rank &  classification based on data availability in key datasets, see Table \ref{table:gsb_data} \\
29 & confirmed & Is it a confirmed GC in NGC~5128? (y/n) \\
30 & references & literature references* \\

\hline
\end{tabular}
\begin{tablenotes}
      \small
      \item The complete table will be available online.
      \item *Reference papers: $^a$\citet{Graham1980},
      $^b$\citet{VanDenBergh1981}, $^c$\citet{Hesser1986}
      $^d$\citet{Harris1992}, $^e$\citet{Holland1999}, $^f$\citet{Harris2002}, $^g$\citet{Peng2004}, $^h$\citet{Martini2004}, $^i$\citet{Harris2004}, $^j$\citet{Woodley2005}, $^k$\citet{Gomez2006}, $^l$\citet{Harris2006},  $^m$\citet{Rejkuba2007}, $^n$\citet{Woodley2007}, $^o$\citet{Beasley2008},  $^p$\citet{Georgiev2009}
      $^q$\citet{Mouhcine2010}, $^r$\citet{Woodley2010b}, $^s$\citet{Woodley2010a}, $^t$\citet{Sinnott2010}, $^u$\citet{Taylor2010}, $^v$\citet{Georgiev2010}, $^w$\citet{Harris2012},
      $^x$\citet{Taylor2015}, $^y$\citet{Taylor2017}, $^z$\citet{Voggel2020},
      $^{aa}$\citet{Fahrion2020}, $^{ab}$\citet{Muller2020}

\end{tablenotes}
\label{table:cands_descr}
\end{table*}

\begin{table*}[ht]
\centering
\caption{Header information for the catalog of all PISCeS sources in NGC~5128}
\begin{tabular}{c c c}
\hline \hline

Column & Label & Description \\
\hline
1 & ID & unique `H21' ID  \\
2 & Discovery\_ID & Discovery ID, if available  \\
3 & RA & PISCeS R.A. in decimal degrees (J2000)  \\
4 & Dec & PISCeS Dec in decimal degrees (J2000)  \\
5 & PISCeS\_gmag & PISCeS $g$ band magnitude (mag) \\
6 & PISCeS\_gmag\_err & error of PISCeS $g$ band magnitude (mag)  \\
7 & PISCeS\_rmag & PISCeS $r$ band magnitude (mag) \\
8 & PISCeS\_rmag\_err & error of PISCeS $r$ band magnitude (mag)  \\
9 & g\_C\_3\_6 & PISCeS $g$ band Concentration Index value  \\
10 & r\_C\_3\_6 & PISCeS $r$ band Concentration Index value  \\
11 & band\_used & PISCeS magnitude band used to calculate \textit{concentration\_likeliness} \\
12 & Gaia\_RA & Gaia DR2 R.A. in decimal degrees (ICRS)  \\
13 & Gaia\_Dec & Gaia DR2 Dec in decimal degrees (ICRS)  \\
14 & Gaia\_Gmag & Gaia DR2 $G$ band magnitude (mag) \\
15 & pm\_ra & Gaia DR2 proper motion measurement in R.A. direction (mas/yr) \\
16 & pm\_ra\_err & error of Gaia DR2 proper motion measurement in R.A. direction (mas/yr) \\
17 & pm\_dec & Gaia DR2 proper motion measurement in Dec direction (mas/yr) \\
18 & pm\_dec\_err & error of Gaia DR2 proper motion measurement in Dec direction (mas/yr) \\
19 & parallax & Gaia DR2 parallax measurement (mas) \\
20 & parallax\_err & error in Gaia DR2 parallax measurement (mas) \\
21 & AEN & Gaia DR2 Astrometric Excess Noise (mas) \\
22 & BR\_excess & Gaia DR2 BP/RP Excess Factor  \\
23 & NSC\_umag & NSC $u$ band magnitude (mag) \\
24 & NSC\_umag\_err & error of NSC $u$ band magnitude (mag) \\
25 & NSC\_gmag & NSC $g$ band magnitude (mag) \\
26 & NSC\_gmag\_err & error of NSC $g$ band magnitude (mag) \\
27 & NSC\_rmag & NSC $r$ band magnitude (mag) \\
28 & NSC\_rmag\_err & error of NSC $r$ band magnitude (mag) \\
29 & NSC\_imag & NSC $i$ band magnitude (mag) \\
30 & NSC\_imag\_err & error of NSC $i$ band magnitude (mag) \\
31 & NSC\_zmag & NSC $z$ band magnitude (mag) \\
32 & NSC\_zmag\_err & error of NSC $z$ band magnitude (mag) \\
33 & ur\_color & NSC $(u - r)_0$, with Milky Way dust correction applied (mag)  \\
34 & rz\_color & NSC $(r - z)_0$, with Milky Way dust correction applied (mag) \\
35 & concentration\_likelihood & candidate likelihood based on $C_{3-6}$ \\
36 & structure\_check & galaxy/blend check \\
37 & proper\_motion\_likelihood & candidate likelihood based on proper motion \\
38 & parallax\_likelihood & candidate likelihood based on parallax \\
39 & aen\_likelihood & candidate likelihood basd on AEN \\
40 & bre\_likelihood & candidate likelihood based on BR$_{\rm excess}$ \\
41 & color\_likelihood & candidate likelihood based on NSC color \\
42 & total\_likelihood & total likelihood of candidate being a GC in NGC~5128 \\
43 & candidate\_rank &  classification based on data availability in key datasets, see \S \ref{table:gsb_data} \\
44 & confirmed & Is it a confirmed GC in NGC~5128? (y/n) \\
45 & references & literature references* \\

\hline
\end{tabular}
\begin{tablenotes}
      \small
      \item The complete table will be available online.
      \item *Reference papers: See Table \ref{table:cands_descr}

\end{tablenotes}
\label{table:all_descr}
\end{table*}

\begin{table*}[ht]
\centering
\begin{tabular}{c c c c c c c }
\hline \hline
Candidate Rank & PISCeS & {\it Gaia} DR2 & NSC & Minimum $C_{3-6}$ & \multicolumn{2}{c}{Number of Candidates} \\
 &        &          &     &                   & Total & $\mathscr{L}_{total} > 0.85$ \\
\hline
gold   & $\checkmark$ & $\checkmark$ & $\checkmark$ & 1.0 & 5763 &  181 \\ 
\hline
silver & $\checkmark$ & $\checkmark$ & $\times$     & 1.0 & \multirow{2}{*}{13793} &  \multirow{2}{*}{1750} \\
       & $\checkmark$ & $\times$     & $\checkmark$ & 1.0 & \\
\hline
bronze & $\checkmark$ & $\times$     & $\times$     & 2.0 & 11860 & 11860 \\
\hline
copper & $\checkmark$ & $\times$     & $\times$     & 1.0 & 9086 & 8541 \\
\hline
\end{tabular}
\caption{In addition to providing a \textit{total\_likelihood}, we identify GC candidates by a rank of `gold', `silver', `bronze', or `copper' based on the amount of data available for each, and columns 2 -- 4 indicate if data is available ($\checkmark$) or not ($\times$) in the three key datasets. 
For many candidates, there is only one source of information (PISCeS) available, and so we label those with $C_{3-6} \geq$ 2.0 as bronze, and those with 1.0 $\leq$ $C_{3-6}$ $<$ 2.0 
as copper, where $C_{3-6} =$ 2.0 corresponds to \textit{concentration\_likelihood} = 0.977 (see \S \ref{sub:2_ap_tech}).
For each rank, the final two columns list the total number of candidates and the number of candidates with \textit{total\_likelihood} $>$ 0.85.  
}
\label{table:gsb_data}
\end{table*}

The first presented catalog includes the 40,502 GC candidates in NGC~5128 found through our GC candidate selection process, as detailed in Section \ref{sec:GCselection}. 
We consider a PISCeS source to be a viable GC candidate if it has $C_{3-6} > 1.0$ and survives applicable yes/no cuts (initial cuts, \textit{structure\_likelihood}, \textit{proper\_motion\_likelihood}, \textit{parallax\_likelihood}; see \S \ref{sec:initial}, \ref{sec:galcul}, \ref{sec:gaia_dr2}). 
The likelihood values for each candidate are multiplied together, and a reported \textit{total\_likelihood} value closer to 1.0 represents a GC candidate with a larger chance of being a true GC in NGC~5128.

Table \ref{table:cands_descr} lists the header information for the catalog of GC candidates, which includes:
(1) a unique `H21' ID, which we assign; (2) Discovery ID; (3, 4) PISCeS R.A. \& Dec.~in J2000 coordinates; (5 -- 8) PISCeS $g$ \& $r$ magnitudes and errors; (9) concentration index ($C_{3-6}$); (10) {\it Gaia} DR2 $G$ magnitude; (11 -- 18) proper motion in the R.A. and Dec. directions, parallax, AEN, and $BR_{\rm excess}$ from {\it Gaia} DR2; and (19, 20)  NSC $(u-r)_0$ and $(r-z)_0$ color with Milky Way dust correction applied on a source-by-source basis \citep{Schlafly2011}. 
We also list a breakdown of the likelihood values (21 -- 27) \textit{concentration\_likelihood}, \textit{proper\_motion\_likelihood}, \textit{parallax\_likelihood}, \textit{aen\_likelihood}, \textit{bre\_likelihood}, \textit{color\_likelihood}, and \textit{total\_likelihood}; and (28) gold, silver, bronze, copper rank (see below).  
For completeness, we (29) include and note targets that have previously been confirmed as GCs and (30) list all literature references.
All PISCeS and {\it Gaia} DR2 photometry reported here has not been corrected for Galactic extinction.  
We do not include any GC candidates from the literature that are outside the PISCeS footprint or within 10\arcmin~from the center of NGC~5128. 
The full GC candidate table will be available online in electronic format.

The second catalog includes all PISCeS sources that pass our initial cuts (\S  \ref{sec:initial}).  In addition to GC candidates, this source catalog includes background galaxies and foreground stars that were removed by various steps in our GC candidate selection process.  
Along with the columns listed in the GC candidate table, we include information about the concentration index from both PISCeS $g$ and $r$ bands; \textit{structure\_check}; {\it Gaia} DR2 R.A. \& Dec.~in ICRS coordinates; and un-corrected NSC $u,g,r,i,z$ photometry and errors. The header information is listed in Table \ref{table:all_descr}. The full PISCeS source table will be available online in electronic format.

\subsection{Globular cluster candidate data rank} \label{sub:gsb}

As a guide for future researchers who may use this GC candidate catalog for spectroscopic follow-up, we split our candidates into four categories (`gold', `silver', `bronze', or `copper') based on the number of our key datasets they appear in (see Table~\ref{table:gsb_data}), separate from their quality (i.e. \textit{total\_likelihood}). 

Gold GC candidates have data from PISCeS, {\it Gaia} DR2, and NSC. 
This group of GC candidates is expected to have the least amount of contamination from foreground stars and background galaxies at high \textit{total\_likelihood} scores, because they have information from all three key datasets pointing to the same result. The trade-off is that the gold group does not include faint GCs, as illustrated by the right panel of Figure \ref{fig:recovery}.  This is primarily due to the lack of {\it Gaia} DR2 sources at $G$ $\gtrsim$ 20 mag, which is $\sim$ 0.5 mag brighter than the NGC~5128 peak of the GCLF \citep{Pota15}.
Of the confirmed GCs outside of 10' (see Appendix \ref{app:confirmed}), 42\% have information in all three datasets. 
There are 5,763 GC candidates in the gold category, of which 181 have \textit{total\_likelihood} $\geq$ 0.85, which we identify as a good threshold in Section~\ref{sec:spatial_dist} because it maximizes the number of GC candidates above the background. 
The gold sample displays a clear overdensity of candidates near NGC~5128, which we discuss further below. Of the gold sample candidates between 10--30 arcmin (32.4 kpc), 15 have \textit{total\_likelihood} $\geq$ 0.85 and have not been been observed spectroscopically.  This suggests that more GCs will be confirmed by following up this sample both close to and far from the center of the galaxy.

Silver GC candidates have data from PISCeS and one of either {\it Gaia} DR2 \emph{or} NSC.  
While the contamination in this set of candidates will be higher than the gold sample, it will also be more sensitive to faint GC candidates.  
Of the confirmed GCs outside of 10', 38\% have information in two datasets. 
The silver sample has a total of 13,793 candidates,  of which 1,750 have \textit{total\_likelihood} $\geq$ 0.85. Between 10--30\arcmin~from the center of the galaxy we find 126 silver GC candidates with \textit{total\_likelihood} $\geq$ 0.85 that have yet to have spectroscopic follow-up or confirmation.

Bronze and copper GC candidates have data only from PISCeS, and are absent from both {\it Gaia} DR2 and NSC.  The likelihood of these candidates is therefore only based on our concentration index measurement $C_{3-6}$ and passing our extended galaxy checks (\S \ref{sub:2_ap_tech}, \ref{sec:galcul}).  Because there is information available from only one catalog, we raise the minimum $C_{3-6}$ value for the bronze candidates to 2.0, corresponding to a \textit{concentration\_likelihood} $>$ 0.977.  The bronze sample contains many promising candidates, of which there are 11,860 in total.  The remaining 9,086 candidates with 1.0 $ \leq C_{3-6} < $ 2.0 are placed in the copper sample. These candidates should be treated with caution, but may contain true NGC~5128 GCs. For reference, 14\% of confirmed GCs outside of 10' have information only in the PISCeS dataset.
These groups of candidates are expected to have the most contamination, and are somewhat susceptible to image-level issues apparent in a handful of fields.  For instance, a changing FWHM across a Megacam field of view can systematically change the average concentration index value at a small level.


\section{Discussion} \label{sec:discussion}

\subsection{Globular cluster spatial distributions} \label{sec:spatial_dist}

\begin{figure*}
\centering
\includegraphics[width=0.48\linewidth]{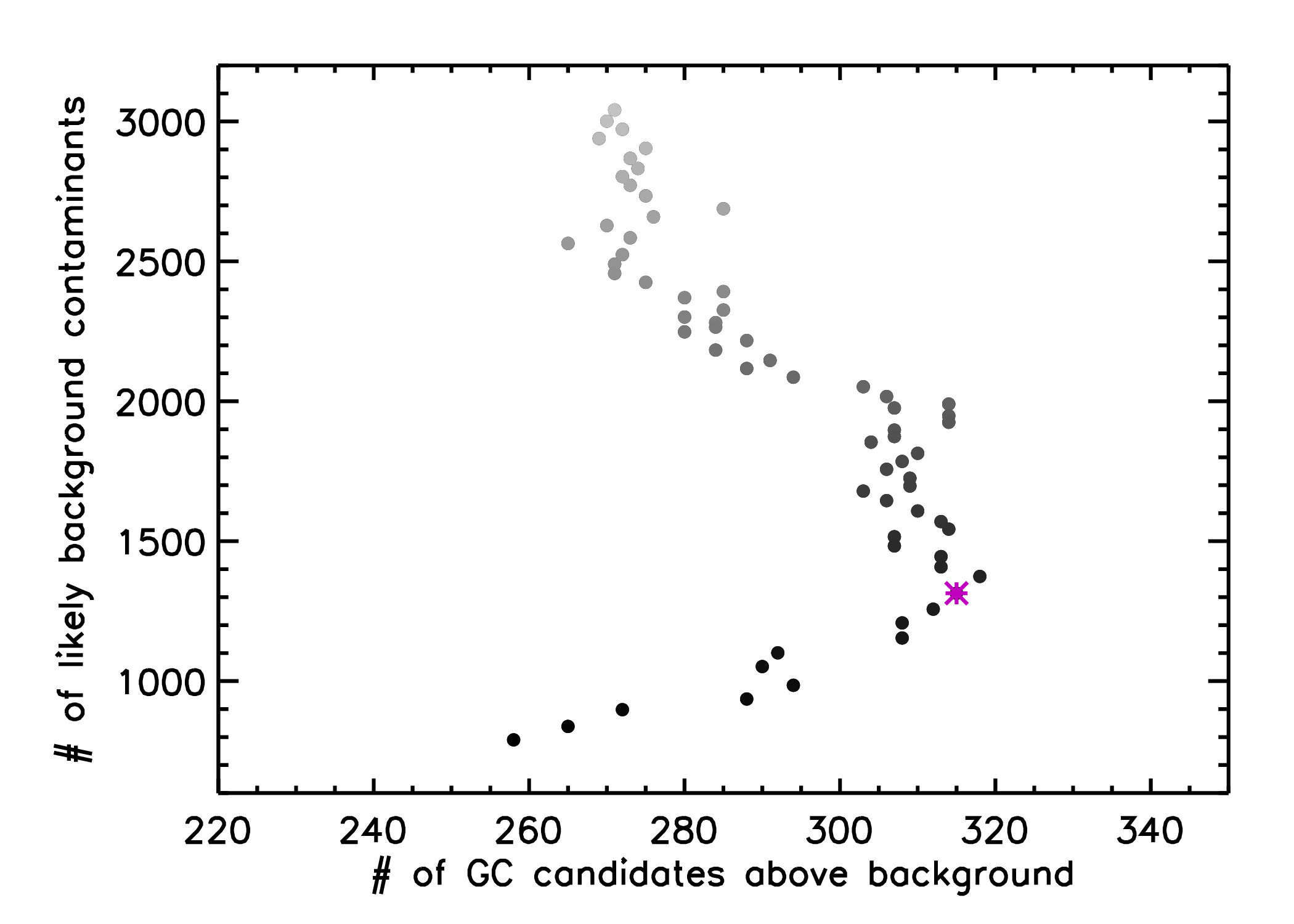}
~
\includegraphics[width=0.48\linewidth]{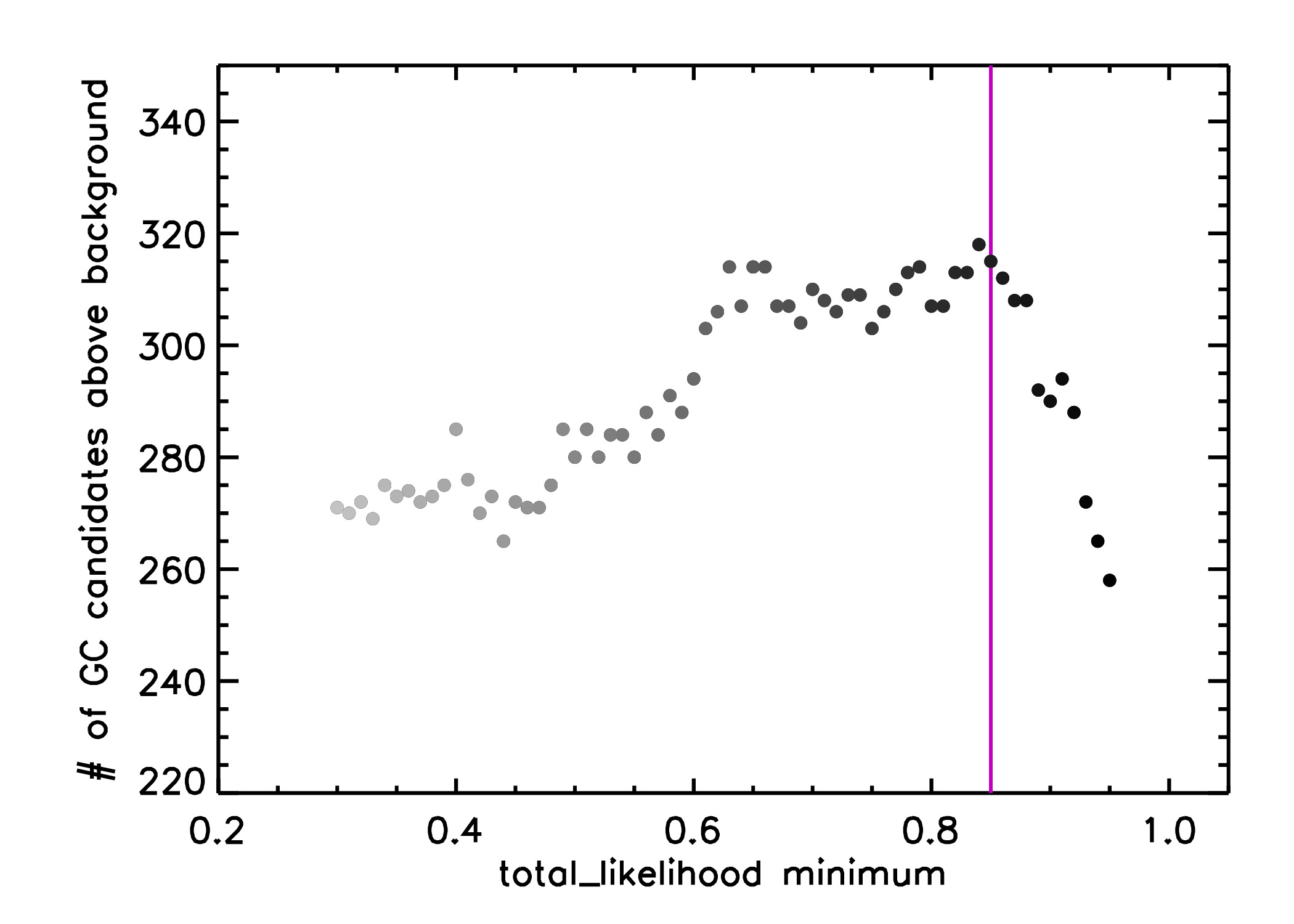}

\caption{Left: The relationship between the number of likely background contaminants and the number of GC candidates above the background level in the gold and silver GC candidate samples with different minimum values of \textit{total\_likelihood} between 0.3 and 0.95. The counts are based on power law fits to the radial distribution of our candidates (e.g. Figure \ref{fig:rad_dist}). Darker colored points correspond to a higher minimum \textit{total\_likelihood} value.  Right: The number of GC candidates above the background as a function of minimum \textit{total\_likelihood}, where the points are color-coded the same as the left panel.
We highlight the gold and silver GC candidates with \textit{total\_likelihood} $\geq$ 0.85 as the most likely to be true GCs based on available information, and this value is indicated with the purple star on the left and the purple line on the right.}
\label{fig:decide_min}
\end{figure*}

Our catalog consists both of true GCs as well as contaminants.  To separate these two, and to optimize our catalogs, we assume that the true GCs follow a decreasing power-law function with radius, while the background level is uniform across our survey.  We therefore fit our radial density profile of GC candidates using a model with a power-law plus uniform background.  We first use this model to determine what an appropriate \textit{total\_likelihood} threshold is for analyzing the physical properties of GCs in our catalog.  
Focusing on the gold and silver GC candidates, we examine the total number of candidates above the background as well as the total number of contaminants at a range of \textit{total\_likelihood} thresholds.  As shown in Fig.~\ref{fig:decide_min}, we find that using a threshold of \textit{total\_likelihood} $\geq$ 0.85 nearly maximizes the number of GC candidates above the background, while minimizing the number of contaminants.  Therefore, for this section we focus our analysis on the gold and silver GC candidates that have \textit{total\_likelihood} $\geq$ 0.85.

\begin{figure}[t]
\includegraphics[width=1.0\linewidth]{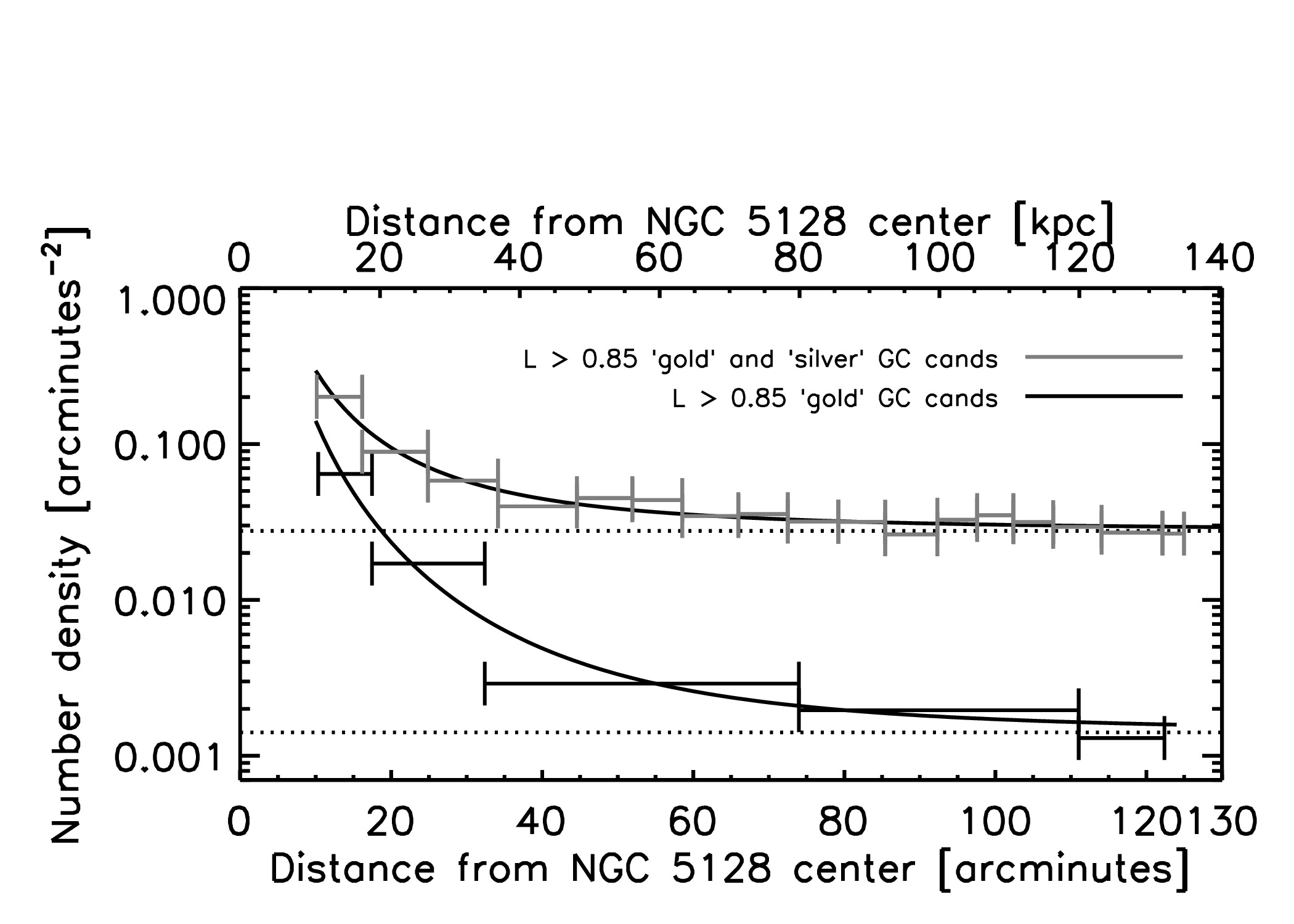}
\caption{ Radial distribution of gold and silver GC candidates with \textit{total\_likelihood} $\geq$ 0.85. Each bin contains 100 GC candidates for the combined gold and silver sample, with 32 entries in the final bin, and 40 GC candidates for the gold-only sample, with 7 entries in the final bin.
The curved solid black lines mark power law fits to the GC candidate distributions between 10\arcmin~and 125\arcmin, given by equations \ref{eqn:density_g} and \ref{eqn:density_gs}, and the dotted lines mark the constant background values in the same equations.}
\label{fig:rad_dist}
\end{figure}

We break down the radial distributions of our combined gold and silver GC candidate population as well as the gold-only GC candidate population in Figure \ref{fig:rad_dist}, where both populations have \textit{total\_likelihood} $\geq$ 0.85. Each radial bin contains a constant number of GC candidates: 100 GC candidates for the combined gold and silver population, with 32 entries in the final bin, and 40 GC candidates for the gold-only population, with 7 entries in the final bin. The power law fit to the radial distribution of the gold GC candidate population is given by
\begin{equation}
    \rho_{g} = 0.00141  + 63.7 r^{-2.66 \pm 0.30}
    \label{eqn:density_g}
\end{equation}
and fit to the distribution of the combined gold and silver GC candidate population is given by
\begin{equation}
    \rho_{g+s} = 0.0277 + 26.6 r^{-1.99 \pm 0.22},
    \label{eqn:density_gs}
\end{equation}
where $\rho_{g}$ and $\rho_{g+s}$ are GC candidate surface densities in N per square arcminute.
These fits are denoted by the curved solid black lines in Figure \ref{fig:rad_dist}.
The combined gold and silver GC candidate population becomes dominated by the background at a smaller radius ($\sim$ 1.4 deg; 94 kpc) than the population with only gold candidates ($\sim$ 1.8 deg; 121 kpc). 

The power-law index from the gold sample ($-2.66\pm0.30$) is comparable to that found for the radial profile of the globular clusters in the Milky Way: \citet{Bica2006} find a 3-D radial power-law index of --3.9 (corresponding to --2.9 in projected radius) for all the clusters beyond 3.5 kpc, and an index --3.6 (--2.6 in projected radius) for the metal-poor clusters that dominate beyond 8 kpc.
These values are steeper than typical massive elliptical galaxies, where the projected power law indices for the full GC systems at large radii are found to range between --1.2 and --2 \citep{Faifer2011}. The index for the combined gold and silver sample ($-1.99\pm0.22$) is consistent with the steeper edge of this range. Overall, the spatial profile appears intermediate between massive galaxies and the typical $L^{\ast}$, Milky Way-like galaxy, though it should be revisited once spectroscopy is available.

As a point of comparison, the SCABS catalog GC candidates within 120\arcmin, excluding an overdensity at a radius of $\sim 55$\arcmin, are fit with a power law of slope --1.22 (T17), shallower than our result. This may partially arise from T17's use of a model that assumes a total population of GCs based on a ``known GC" population that contains contamination from foreground stars (see \S \ref{sec:compare} and Appendix \ref{app:misconfirmed}).

\begin{figure}[t]
\includegraphics[width=1.0\linewidth, trim={0cm 0cm 19cm 0cm},clip]{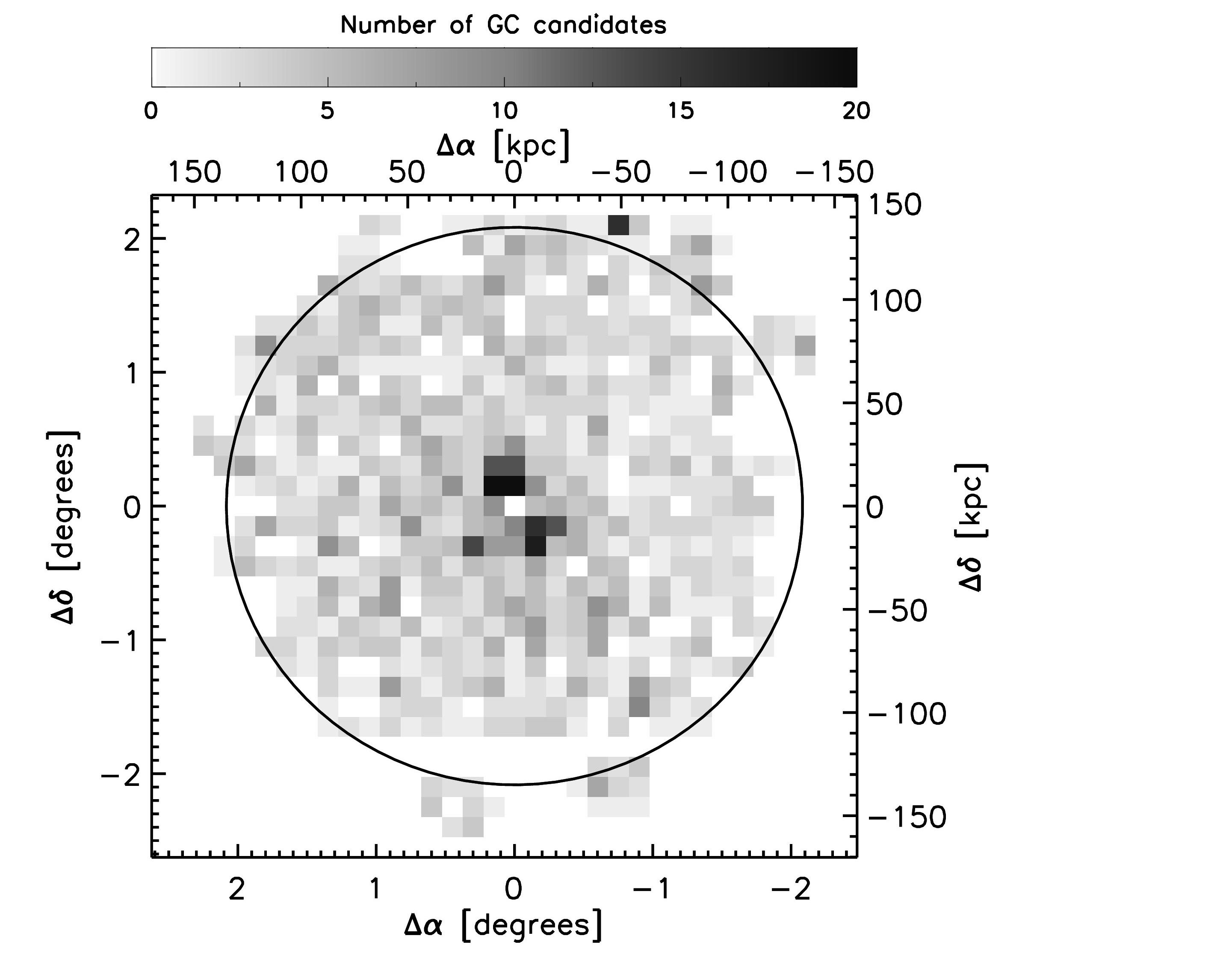}
\caption{The two-dimensional density distribution of our gold and silver GC candidates with a \textit{total\_likelihood} $\geq$ 0.85, where each pixel is 0.15$\times$0.15 degrees.  The figure is oriented such that north is up and east is left. Note the overdensity of GC candidates close to the center of the galaxy, which is angled similarly to the isophotal major axis of the galaxy (35$^\circ$ E of N, \citealt{Dufour1979}). 
We exclude all sources within 10\arcmin from our analysis, leading to a central gap in the GC candidate distribution. The circle marks a distance of 125\arcmin (135 kpc) from the galaxy center, which is about the extent to which PISCeS has full coverage.}
\label{fig:density}
\end{figure}

The two-dimensional distribution of our gold and silver GC candidates with  \textit{total\_likelihood} $\geq$ 0.85 is shown in Figure \ref{fig:density}.  There is a central overdensity, especially along the diagonal from north-east to south-west out to a distance of about 30\arcmin~(32.4 kpc), which is angled similarly to the isophotal major axis of the galaxy (35$^\circ$ E of N, \citealt{Dufour1979}).  
A majority of the already confirmed GCs in NGC~5128 lie within this radius, though the position angle overdensity seen here has a broader distribution than has been found in previous works.  Though we see an overdensity in the one-dimensional GC candidate distribution out to $\sim$ 85\arcmin, it is less apparent in the two-dimensional distribution. While not plotted, we also see a central (within 40--50\arcmin) enhancement of the bronze candidates above background, consistent with at least some of these being real GCs.

\subsection{GC total population estimates} \label{sec:population}

\begin{figure*}
\centering
\includegraphics[width=0.48\linewidth]{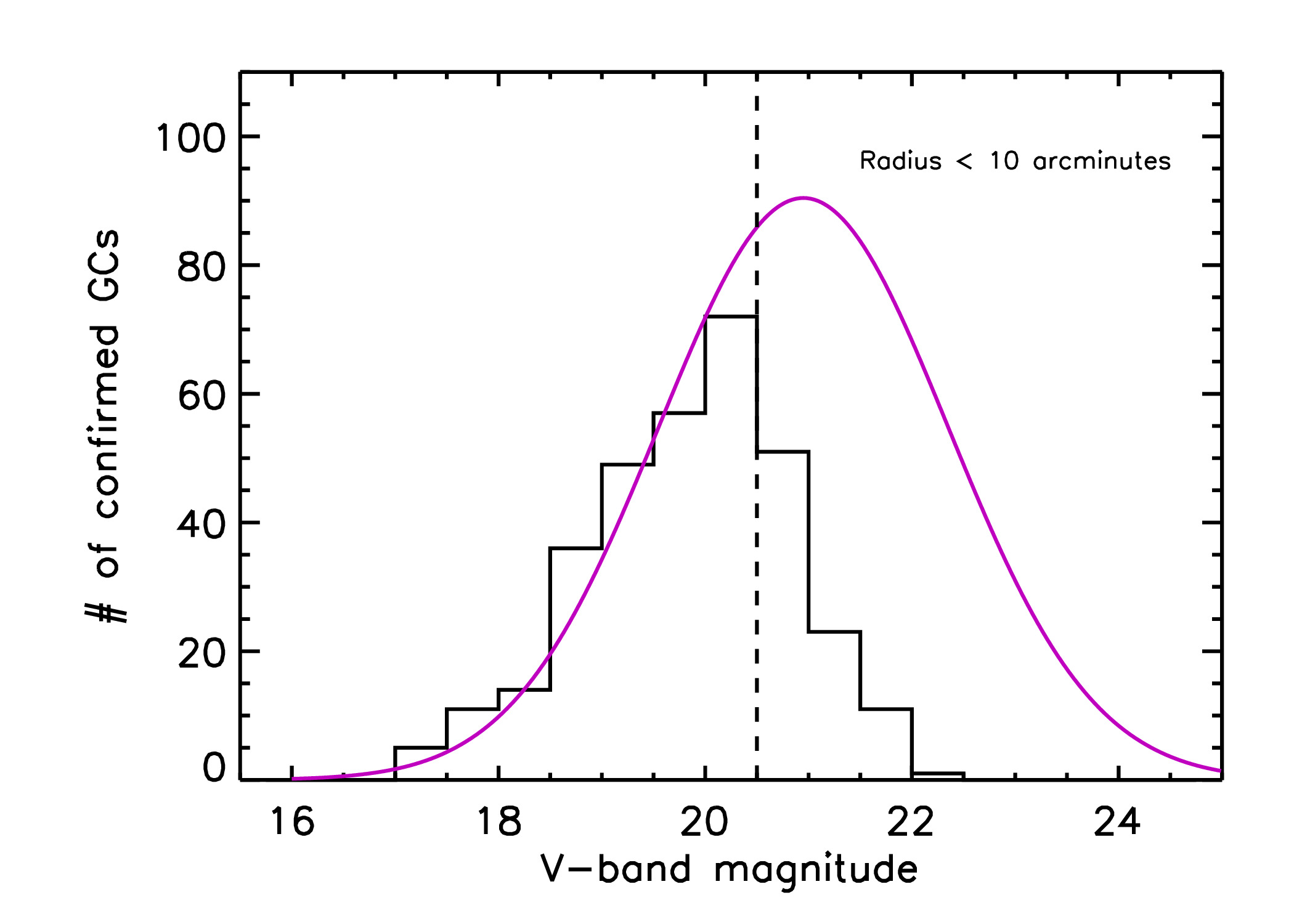}
~
\includegraphics[width=0.48\linewidth]{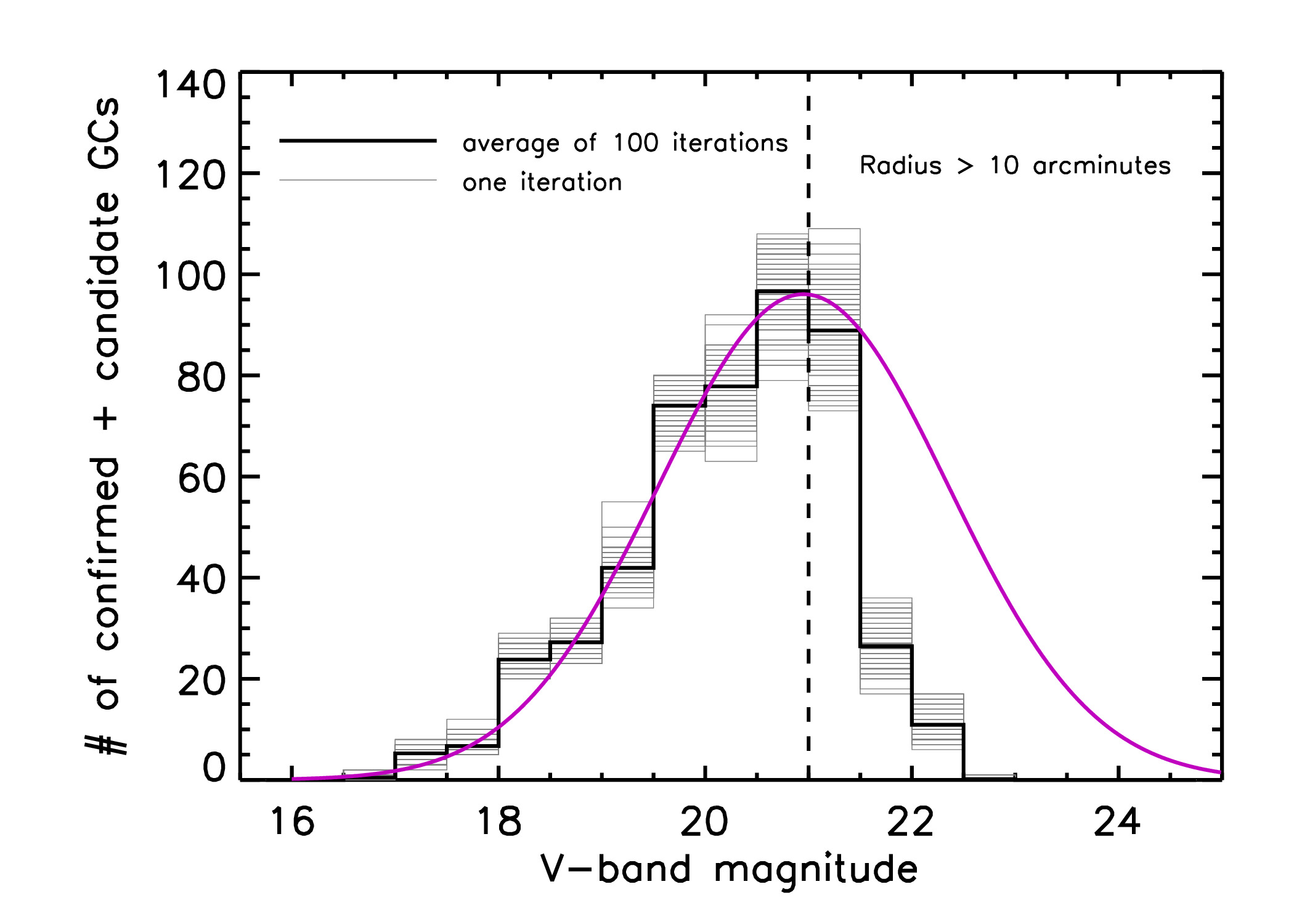}
\caption{An illustration of our process to calculate the total number of GCs in NGC~5128, using Gaussian functions fitted to the expected GCLF, split at a radius of 10\arcmin. The confirmed GC population within 10\arcmin~(left) is assumed to be complete for $V \lesssim 20.5$ mag.  Outside of 10\arcmin (right), the known population of confirmed GCs is combined with a background-subtracted sample of our gold and silver candidate GCs with \textit{total\_likelihood} $\geq$ 0.85.  
Plotted here are 100 iterations of these background corrected candidate sampling (in grey), and their average (in black). Only data to the left of the dashed vertical lines in both histograms are used in the respective Gaussian fits.}
\label{fig:mag_hist}
\end{figure*}

Using a method similar to \citet{Harris2010_alone}, we estimate the total size of the GC population in NGC~5128 by forcing a fit to the standard GCLF and estimating the number of GCs below our detection limit.  
Based on values of $(m - M)_0 = 27.91$ \citep{Harris2010}, $A_V = 0.35$ \citep{Schlafly2011}, and assuming a mean turnover magnitude $M_V = -7.3$ \citep{Harris2001}, 
the GCLF turnover is expected to occur at $V = 20.95$ mag with $\sigma$=1.4 mag.  We convert our PISCeS $g$- and $r$-band magnitudes to $V$-band using transformation coefficients from \cite{Jester2005}.

We split the expected population calculation at a radius of 10\arcmin, and assume that within this radius, all bright ($V \leq 20.5$ mag) GCs have been found.
Outside of 10\arcmin, we combine the known confirmed GCs with a sample of our new gold and silver candidate GCs with \textit{total\_likelihood} $\geq$ 0.85.  
The sample of GCs is chosen randomly from our candidates to correct for background contamination based on the fits shown in Fig.~\ref{fig:rad_dist}.  These random samples were chosen 100 times. 
For both the inner and outer cluster samples we plot the GCLF and fit a Gaussian function with values discussed above assuming Poisson errors.  We also varied the faint end limit of the fitting range by 0.5 magnitudes, and incorporate this into  our overall uncertainties.

This method results in a population estimate of 1300 $\pm$ 110 GCs, which closely matches the total predicted by \citet{Harris2010_alone}.  This should be treated as a lower limit though, since it does not included any GC candidates from the bronze or copper classifications (those that have data only from PISCeS), and it includes only gold and silver clusters with \textit{total\_likelihood} $\geq$ 0.85. We know that 14\% of confirmed GCs outside of 10\arcmin~are in the bronze sample, largely due to the incompleteness of {\it Gaia} DR2 and NSC at faint magnitudes.  Another 4\% of confirmed GCs are in the gold and silver sample but have \textit{total\_likelihood} $<$ 0.85.  
Accounting for this incompleteness, we estimate that there are a total of 1450 $\pm$ 160 GCs in NGC~5128 within $\sim$~150 kpc.

\citet{Forbes2017} predicts that half of the GC system of an elliptical galaxy will lie within a radius $R_{e}$ given by $\log R_e = 0.97 (\pm 0.4) \log M_* - 9.76 (\pm 4.4) $. For a stellar mass of $5.5 \times 10^{11} M_\odot$ \citep{Peng2004}, this result predicts that half of NGC~5128's GCs will lie within a radius of $42^{+42}_{-21}$ kpc.  We find that half of our estimated total GC population lies within a radius of 48 $\pm$ 3 kpc. Only three GCs beyond this radius have been confirmed, further illustrating the need for spectroscopic confirmation of GCs in the outer halo.

The specific frequency ($S_N$) of a galaxy is used to measure the relative occurrence of 
GCs across galaxy types and masses, and is defined as $S_N = N_{GC} \times 10^{0.4(M_V + 15)}$ \citep{Harris1981}.  Using our new GC population estimate of $N_{GC}$ = 1450 and an extinction-corrected V-band magnitude of $M_V$ = --21.38 for NGC~5128 \citep{RC3_paper, Schlafly2011} results in $S_N$ = 4.1. This is somewhat higher than other recent $S_N$ calculations for NGC~5128, including $S_N$ = 2.9 (T17) and 1.9 \citep{Harris2010_alone}. These differences are due both to different $N_{GC}$ estimates and to different assumed $M_V$ values.

Our calculation ($S_N$ = 4.1) is consistent with the idea of \citet{Graham1979} that NGC~5128 experienced a merger between an elliptical galaxy with a relatively high $S_N$, and a second gas-rich galaxy, which would typically have had a lower $S_N$.


\subsection{Comparison with previous results} \label{sec:compare}

\begin{figure*}
\centering
\includegraphics[width=0.48\linewidth]{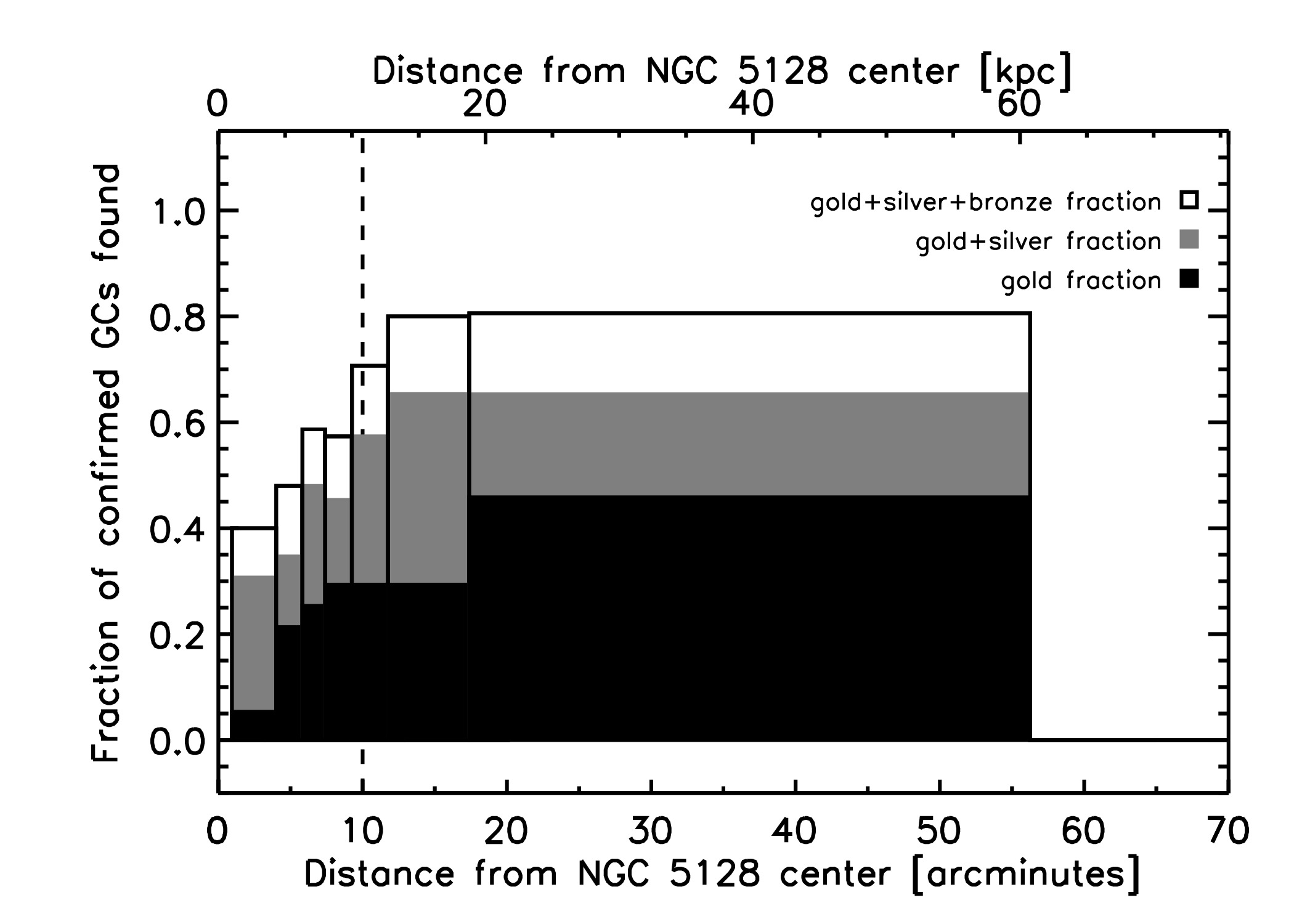}
~
\includegraphics[width=0.48\linewidth]{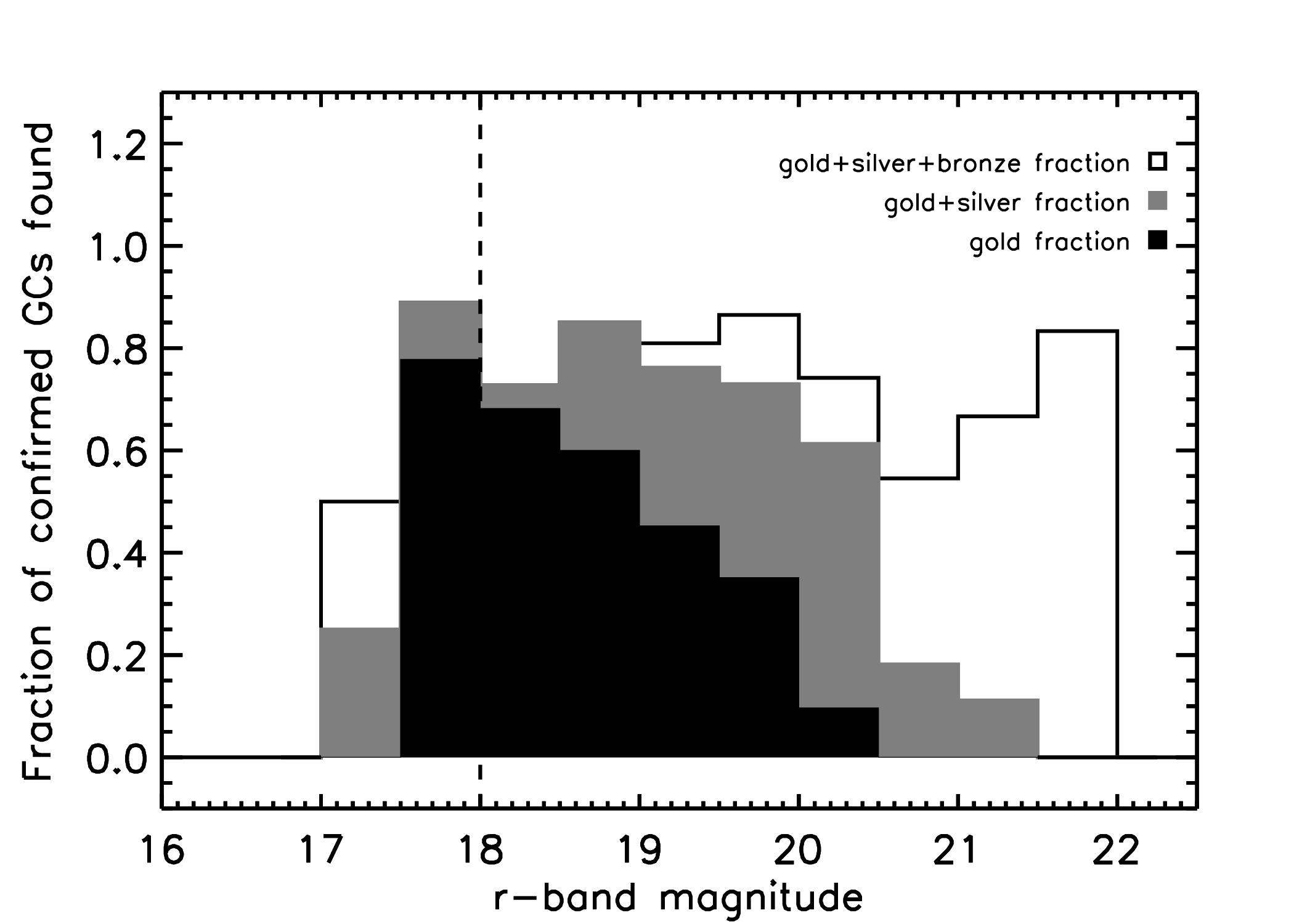}

\caption{Left: Recovery fraction of confirmed GCs with \textit{total\_likelihood} $\geq$ 0.85, binned by sets of 75 with increasing radial distance from the center of NGC~5128. The last bin has 72 GCs.  Within 10\arcmin of the galaxy center our GC selection process is not effective due to crowding.  Right: Recovery fraction with \textit{total\_likelihood} $\geq$ 0.85 of the confirmed GCs outside of 10\arcmin, binned by $r$-band magnitude, without extinction correction applied.  
The dashed vertical line indicates where {\it Gaia} DR2 begins to become incomplete \citep{GaiaMain18}, and thus where we expect the gold sample to decrease.  }
\label{fig:recovery}
\end{figure*}

Cross-matching items in our GC candidate catalog with the fiducial sample of 69 confirmed GCs, we find that 65 are recovered with \textit{total\_likelihood} $\geq$ 0.85  
(this is, of course, largely by construction as we used the fiducial sample to guide our GC candidate selection methodology).  
The remaining four fiducial sample GCs have positions in color-color space near the border of the color selection ellipse discussed in \S \ref{sec:color}, and therefore have lower values for their final \textit{total\_likelihood} values.

We broaden our comparison sample to include all confirmed GCs; see Appendix~\ref{app:confirmed} for a list.  Comparisons of recovery fractions for the gold, silver, and bronze rank classifications as a function of radius and magnitude are shown in Figure \ref{fig:recovery}; no confirmed GCs belong to the copper rank.  The requirement for data from the NSC and {\it Gaia} DR2 catalogs causes faint end completeness to drop in the gold and silver GC candidate populations.
We recover a total of 192 out of 198 (97\%) confirmed GCs that are outside of 10\arcmin~and in PISCeS. 
Additionally, 155 out of 198 (78\%)  have \textit{total\_likelihood} $\geq$ 0.85, where 69 GCs are in the gold sample, 56 GCs are in the silver sample, and 30 GCs are in the bronze sample. 
Those confirmed GCs that are not recovered or have low \textit{total\_likelihood} values  typically have low concentration index values or were identified as galaxies due to close proximity to foreground stars.
Within 10\arcmin, we only recover 53\% of the confirmed GCs, due to crowding. 
Recall that because our methods are tailored to work in the uncrowded regions far from the center of NGC~5128, we exclude the central-most portion of the galaxy from our catalogs.

We fail to recover a small number of bright GCs/UCDs, mostly due to saturation of the PISCeS data, but also because of our cuts on structure that identify background galaxies.  This is why we developed a focused approach in V20 to obtain a complete sample of the brightest GCs, including cluster candidates lying outside the PISCeS footprint. Of the 480 candidates from V20 that lie within the PISCeS footprint, only 102 are also included in our catalog.

Apart from V20, the only other catalog presenting GC candidates at comparably large radii is the SCABS catalog from T17.  We find that 864 of the 2087 GC candidates from SCABS that are in the PISCeS footprint are recovered as candidates by our tests, and 511 have \textit{total\_likelihood} $\geq$ 0.85 in our catalog.  
An additional $\approx$~1100 SCABS GC candidates are outside of the PISCeS footprint or are within 10' of the center of the galaxy and so are not included in our catalog.  
Of the $\approx$ 1450 SCABS candidates with \textit{total\_likelihood} $<$ 0.85 in our catalog, 67\% are found to have stellar-like concentration index values in our PISCeS data and 37\% have high confidence parallax or proper motion measurements in {\it Gaia} DR2.  This suggests that the catalog has a high contamination rate, and that the PISCeS concentration information provides valuable complementary data to the colors used in the SCABS candidate selection. 
Additionally, SCABS recovered only 251 out of their sample of 643 ``confirmed' GCs 
\footnote{The sample of ``confirmed" GCs used by T17 is not the same as what we list in Appendix \ref{app:confirmed}.  T17 does not provide a list of the confirmed GCs that they use to assess their GC candidate selection technique, and cite private communication as a source. Their sample likely includes contamination from foreground stars; see Appendix \ref{app:misconfirmed}. }. 
Before their final step of deriving GC candidate likelihoods based on color, however, they reported that they successfully recovered $\sim$90\% of confirmed GCs. 
This low recovery rate in the final step appears to be caused by their method of modeling the ratio of stars to GCs, which is controlled by an expected population size.
T17 additionally predicts a total of $\sim$~1100 GCs within a projected 55 kpc from the center of NGC~5128, and $\approx$ 3200 GCs within $\sim$ 140 kpc. 
This is more than double our prediction for the total number of GCs in the galaxy (\S \ref{sec:population}).

\subsection{Color bimodality} \label{sec:color_bimodality}

\begin{figure*}
\centering
\includegraphics[width=0.48\linewidth]{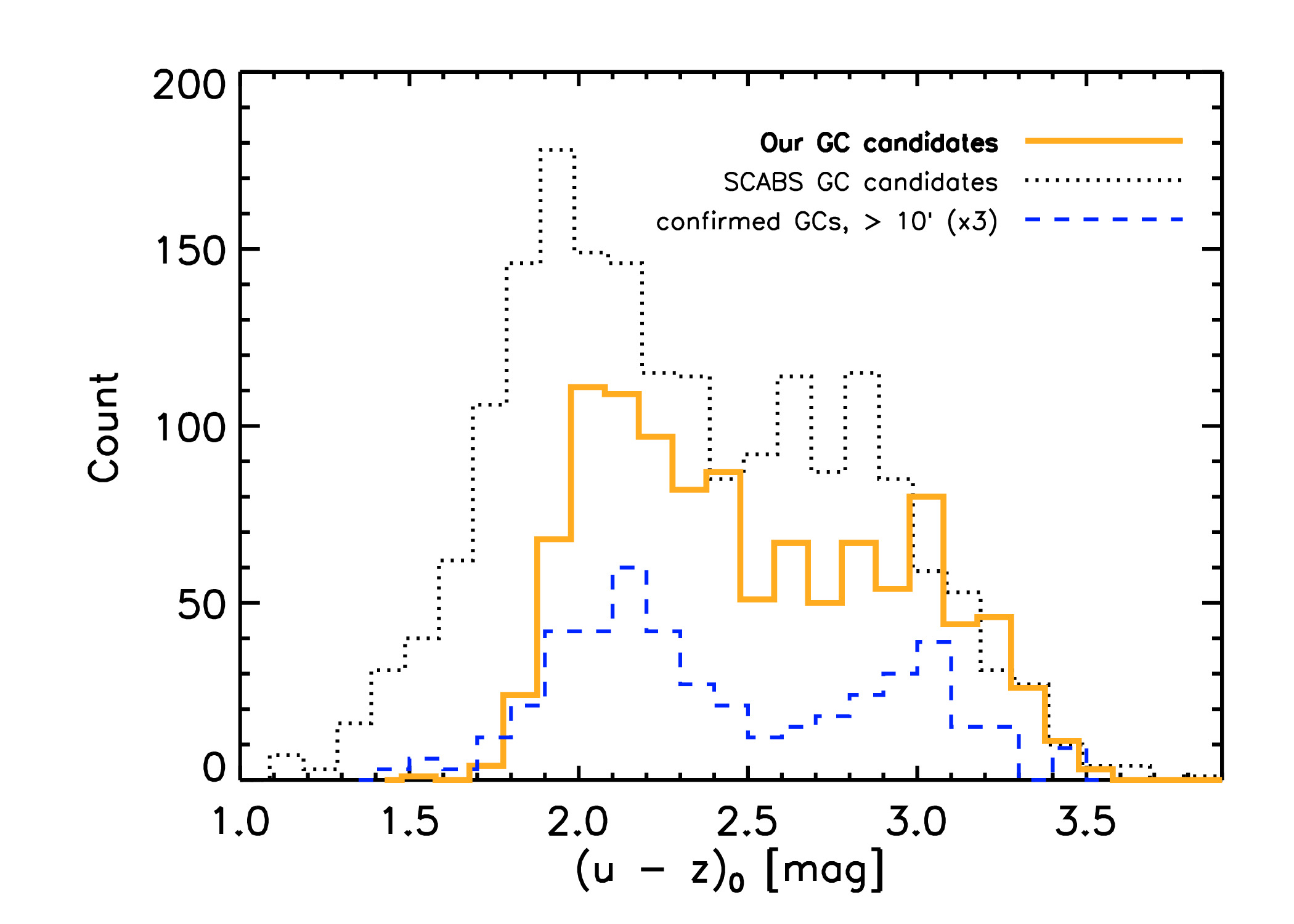}
~
\includegraphics[width=0.48\linewidth]{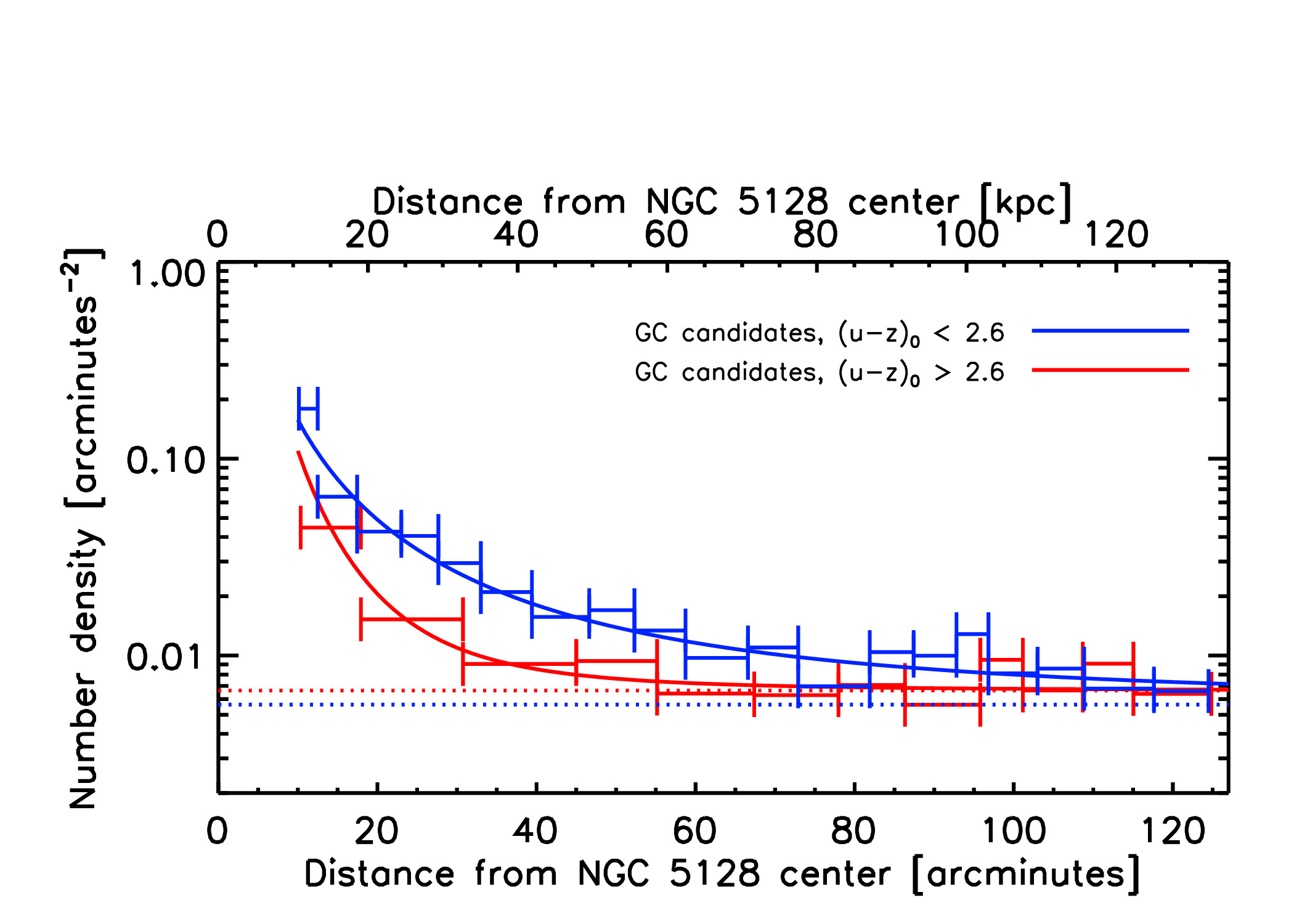}
\caption{
Distributions of gold and silver GC candidates with NSC photometric data and \textit{total\_likelihood} $\geq$ 0.85. 
Left: The bimodal shape of our GC candidate color distribution closely matches that of  the confirmed GCs outside of 10\arcmin~from the center of the galaxy, which is multiplied by 3 to more clearly show its shape. There is an offset between the color of our GC candidates and those found with the methods employed by T17.  NSC photometry has been dust corrected.
Right: Radial distribution of GC candidates, split based on color with a critical value of $(u-z)_0 = 2.6$ mag.  Each radial bin contains 30 GC candidates, with 20 in the final blue bin and 28 in the final red bin.  The curved solid lines mark the power law fits to the GC candidate distributions between 10 and 125\arcmin, given by Equations \ref{eqn:density_blue} and \ref{eqn:density_red}, and the dotted lines mark the constant background values in the same equations. }
\label{fig:color_hist}
\end{figure*}

In many galaxies, the population of GCs is observed to have a distinctly bimodal metallicity distribution: 
though most GCs are metal-poor, the ``blue" GCs in a given galaxy are on average more metal-poor and the ``red" GCs are on average more metal-rich
(e.g., \citealt{Brodie06} for a review). The difference may be the result of dual halo mode formation, where metal-rich GCs form in massive progenitors and metal-poor GCs form in low mass satellite galaxies that are later accreted as part of hierarchical galaxy formation (e.g., \citealt{Park2013, Lee2016, Forbes2018, 2019MNRAS.482.4528E}). This bimodality is one of the most studied properties of GC systems, since these subpopulations show differences in their intrinsic characteristics, such as their spatial distribution and kinematics.

As expected, there is a bimodal color distribution of our gold and silver GC candidates with \textit{total\_likelihood} $\geq$ 0.85.  
Figure \ref{fig:color_hist} shows the distribution in NSC $(u - z)_0$ color space of these candidates. 
The right and left edges of our candidate color distribution matches the distribution of the confirmed GCs that have NSC data and are outside of 10\arcmin, which has been multiplied by a factor of three to show the distribution's shape.  
This match between the color ranges
of our candidates and the confirmed GCs is largely by construction, as we used a subset of confirmed GCs to guide the color selection step in our GC candidate selection process (\S \ref{sec:color}).
 
We also plot the NSC color of the SCABS GC candidates from T17.
The offset in the color distributions between the SCABS GC candidates and the confirmed GCs may arise because of the high rate of contamination in the SCABS sample, as discussed in \S \ref{sec:compare}.

The radial distributions of the same gold and silver GC candidates are plotted in Figure \ref{fig:color_hist}, where GC candidates with $(u-z)_0 < 2.6$ are in the blue sample and GC candidates with $(u-z)_0 > 2.6$ are in the red sample.
The power law fit to the radial distribution of the blue GC candidates is given by
\begin{equation}
    \rho_{b} = 0.00562  + 9.72 r^{-1.81 \pm 0.23}
    \label{eqn:density_blue}
\end{equation}
and the fit to the radial distribution of the red GC candidates is given by
\begin{equation}
    \rho_{r} = 0.00664 + 79.3 r^{-2.89 \pm 0.64}.
    \label{eqn:density_red}
\end{equation}
The slope obtained for the red subpopulation is steeper than the one obtained for the blue subpopulation and the one for the entire gold and silver GC candidate system (Eq. \ref{eqn:density_gs}), as it is more concentrated towards the center.  This follows what has been observed for many galaxies (e.g., \citealt{Brodie06, Faifer2011, Forbes2012}). At all radii that are include in the fit, there are more blue than red GCs, further indicating that NGC~5128 has a rich merger history
(e.g., \citealt{Peng2004, Beasley2008}).


\subsection{Milky Way foreground star contamination}\label{sec:contamination}

Because of the relative proximity of NGC~5128 to the galactic disk there is significant contamination from foreground stars.  We find that nearly all of the objects that lie within the magnitude range expected for NGC~5128 GCs are actually Galactic foreground stars.  Unfortunately, removing all the foreground star contamination from GC catalogs is very tricky. 
Of the 630 GCs identified in previous studies, 355 have both a measured radial velocity and information in {\it Gaia} DR2.  
A histogram of the radial velocities of these GCs is plotted in Figure \ref{fig:rv_hist}.  Overlaid in blue, we show the 50 objects that have non-zero proper motion and/or parallax measurements with a confidence of 3$\sigma$ or greater.  
These objects are contaminant foreground stars with high velocities, not GCs in NGC~5128 as previously thought, and are listed and discussed further in Appendix \ref{app:misconfirmed}. 

We estimate that up to $\approx$80\% of entries with gold or silver rank and \textit{total\_likelihood} $\geq$ 0.85, and $\approx$42\% of entries with gold rank and \textit{total\_likelihood} $\geq$ 0.85, are not true GCs in NGC~5128, based upon the density of the background in the fit to the radial distribution of candidates, shown in Figure \ref{fig:rad_dist}.  We expect this percentage to be dependent on the radial distance from the center of the galaxy and the amount of information available for the candidates.

\subsection{Velocity Dispersion of GCs Around NGC~5128}

\begin{figure}[t]
\includegraphics[width=1.0\linewidth]{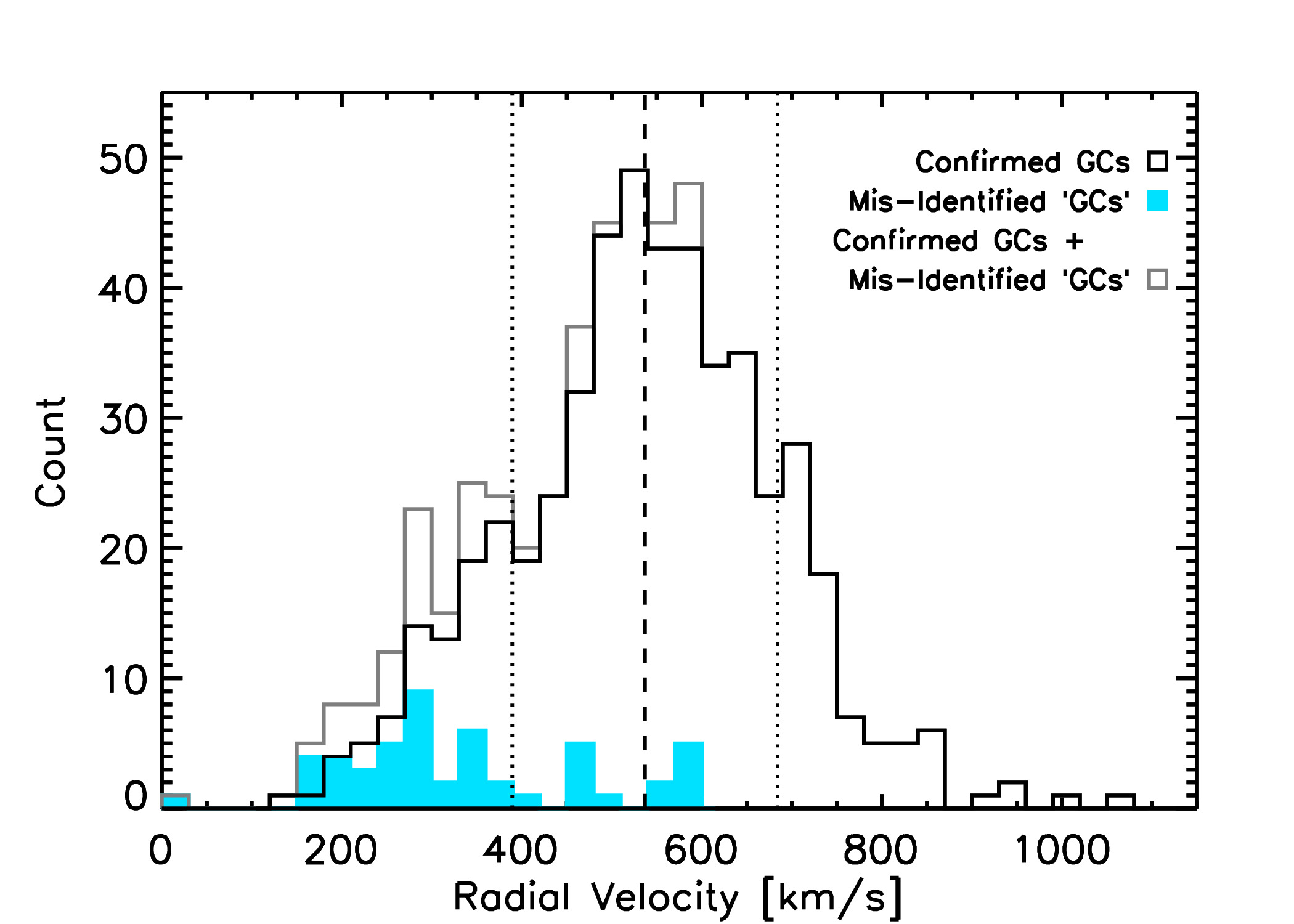}
\caption{ Histogram of 507 confirmed GCs with radial velocity measurements.  The dashed line marks the newly measured mean radial velocity value of 537 km s$^{-1}$ and the dotted lines mark $\pm 1 \sigma$, where $\sigma$ = 148 km s$^{-1}$.
The standard critical radial velocity for GC confirmation in NGC~5128 is 250 km s$^{-1}$. Targets with velocities below this value were typically confirmed via structural parameters or visual resolution into component stars from {\it HST} images.
The filled blue histogram shows the 50 objects that were identified by a past NGC~5128 survey as `GCs' that have non-zero proper motion measurements in {\it Gaia} DR2 of 3$\sigma$ or greater. These mis-identified `GCs' are actually Milky Way foreground stars, and are listed and discussed further in Appendix \ref{app:hvfs}. By removing these contaminants, the distribution of radial velocities becomes more symmetric about the mean value. }
\label{fig:rv_hist}
\end{figure}

The systematic velocity of NGC~5128 is 541 km s$^{-1}$ \citep{Hui1995} with a velocity dispersion of $\sigma \sim 150$ km s$^{-1}$ \citep{Wilkinson1986,Silge05}.
Based on a sample of 605 identified GCs, 
\citet{Woodley2010a} calculated a mean GC radial velocity of $517 \pm 7$ km s$^{-1}$ with $\sigma = 160 \pm 5$ km s$^{-1}$.  They note that because this value is lower than the accepted systematic velocity of the galaxy, their catalog may be contaminated on the low velocity end.  We have found and removed contaminant foreground stars from the confirmed GC population (see Appendix \ref{app:misconfirmed}), and have redone the calculation. 
Of the cleaned list of confirmed GCs (see Table \ref{table:confirmed_short} in the Appendix), 507 have radial velocity measurements, 
with a mean radial velocity of 537 km s$^{-1}$ and $\sigma$ = 148 km s$^{-1}$.  
Figure \ref{fig:rv_hist} shows radial velocity histograms for the cleaned list of confirmed GCs, the contaminant Milky Way foreground stars that were previously mis-identitified as GCs, and the combination of these two populations.
Removing the contaminant foreground stars makes the distribution of the GC radial velocity measurements more symmetric about the mean value and brings the mean velocity measurement of the GC system closer in line with that of the galaxy as a whole, as is seen in other galaxies (e.g. \citealt{Pota15}). We defer a thorough kinematic and dynamical analysis to a future paper that will also incorporate new radial velocities.

\subsection{Future work}

Now that we have a new catalog of NGC~5128 candidate GCs out to $\sim$150 kpc, further observations are required to confirm which are true GCs in NGC~5128.  
We have begun taking spectroscopic observations to measure candidate GC radial velocities using Magellan/Megacam, AAT/2DF, and SOAR.  
For brighter GCs we will also be able to estimate metallicities, providing further information on NGC~5128’s accreting substructures. 
The results will be published in an upcoming paper.

Once we have a larger population of confirmed GCs at large radii we will be able to identify specific GCs that are associated with streams and shells around NGC~5128, similar to the work done for M31 \citep{Veljanoski2014, Mackey2019}.  The kinematics of GCs associated with NGC~5128's streams and shells will provide important constraints on the accreted systems' orbit, mass, and concentration while the velocity dispersion, number, and color of the non-associated GCs can provide complementary constraints on the host galaxy mass.  It is also possible that many of NGC~5128's past merger/accretion events may not have visible remnants in the PISCeS RGB map, but could be detectable kinematically via its GCs.  Additionally, using GCs at larger radii we will be able to calculate the galaxy's enclosed mass at a larger radius.

The release of {\it Gaia} Early Data Release 3 and the upcoming full {\it Gaia} Data Release 3 include information on additional targets around NGC~5128, to a fainter limiting magnitude of $G \approx 21$ and with higher precision proper motion and parallax measurements than {\it Gaia} DR2 \citep{Gaia_edr3}.  This will allow for the removal of additional contaminant foreground stars from the silver, bronze, and copper GC candidate categories and a more confident \textit{total\_likelihood} score that takes into account structural parameters (ex. AEN and $BR_{\rm excess}$) given to those candidates that now have {\it Gaia} data available for them.


\section{Summary}\label{sec:conclude}

We present a new technique for identifying GCs in the Local Volume using a combination of ground- and space-based observations.  
We combine data from {\it Gaia} Data Release 2 with the the ground-based PISCeS and NOAO Source Catalog datasets to select 40,502 GC candidates in NGC~5128 out to a projected radius of $\sim$ 150 kpc, as presented in Table \ref{table:cands_descr}.  We highlight the 1,931 candidates in the gold and silver rank with \textit{total\_likelihood} $\geq$ 0.85 as the most likely to be true GCs in NGC~5128, and therefore the most promising targets for followup spectroscopic confirmation.  

Our GC candidate selection process identifies extended objects using a concentration index vs. magnitude relationship, and then rejects foreground star and background galaxy contaminants based on astrometric motion and structural parameters from {\it Gaia} DR2, and color from NSC.  
To assist in selecting extended objects, we provide 3$\sigma$ confidence lines in {\it Gaia} AEN and BR$_{\rm excess}$ vs. $G$-mag space, and a color selection ellipse in NSC ($u-r$)$_0$ vs. ($r-z$)$_0$ space.  
Each GC candidate in our catalog has a \textit{total\_likelihood} value between 0 and 1 (where numbers closer to 1 represent a larger chance that the source is a GC) as well as a gold, silver, bronze, or copper rank (to indicate the amount of information available). Our selection technique recovers 155 out of 198 (78\%) confirmed GCs that are outside of 10\arcmin~and in PISCeS with a \textit{total\_likelihood} $\geq$ 0.85.  

Our sample of gold and silver GC candidates exhibit a clear central overdensity.  The one-dimensional radial distributions of the gold and combined gold and silver GC candidate populations can be fit by power laws with exponents of $-$2.66 and $-$2.01, respectively.   Using expected parameters to fit a GCLF to the combined confirmed and candidate samples, we estimate that there are a total of 1450 $\pm$ 160 GCs in NGC~5128 within $\sim$~150 kpc, and that half of the GCs lie within a radius of 48 $\pm$ 3 kpc.

Our GC candidates exhibit a bimodal color distribution matching that of the confirmed GCs outside of 10\arcmin.  The metal-rich subpopulation is more centrally concentrated, with a power law exponent of $\sim -2.9$, though the metal-poor subpopulation, with a power law exponent of $\sim -1.8$, is larger at almost all radii.

We cull contaminant objects from the list of known confirmed GCs and provide an updated mean radial velocity for confirmed GCs of 537 km s$^{-1}$ with $\sigma = 148$ km s$^{-1}$, which is closer in line with the values for the galaxy as a whole.  The cleaned list of confirmed GCs can be found in Table \ref{table:confirmed_short}.

Once GCs at large radii have been confirmed, we will be able to identify those that are associated with streams and shells around NGC~5128, providing information about the accreted systems and host galaxy, as has been done in M31 \citep{Veljanoski2014, Mackey2019}.
We will additionally apply these techniques combining {\it Gaia} with ground-based observing to identify slightly extended GCs in further galaxies in the Local Volume. 


\acknowledgments

This research uses services or data provided by the NOAO Data Lab. NOAO is operated by the Association of Universities for Research in Astronomy (AURA), Inc. under a cooperative agreement with the National Science Foundation.

This work has made use of data from the European Space Agency (ESA) mission
{\it Gaia} (\url{https://www.cosmos.esa.int/gaia}), processed by the {\it Gaia}
Data Processing and Analysis Consortium (DPAC,
\url{https://www.cosmos.esa.int/web/gaia/dpac/consortium}). Funding for the DPAC
has been provided by national institutions, in particular the institutions
participating in the {\it Gaia} Multilateral Agreement.

AKH and DJS acknowledge support from NSF grants AST-1821967 and 1813708. ACS acknowledges support from NSF grant AST-1813609. JS acknowledges support from NSF grants AST-1514763 and AST-1812856 and from the Packard Foundation. JDS acknowledges support from NSF grant AST-1412792. Research by DC is supported by NSF grant AST-1814208.

We thank Pauline Barmby for supplying globular cluster structural properties from \citet{McLaughlin2008}.

Lastly, we thank the anonymous reviewer whose comments and suggestions helped improve and clarify this manuscript.

%

\vspace{5mm}
\facilities{Magellan:Clay (Megacam), {\it Gaia}}


\software{Source Extractor \citep{Bertin96},  SWarp \citep{Bertin2010}, The IDL Astronomy User's Library \citep{IDLforever}
          }

\bibliographystyle{apj}
\bibliography{thepaper}

\appendix


Based upon new observations and data, we can show that only 557 of the 630 objects that have been identified as NGC~5128 globular clusters (GCs) in the literature  are genuine confirmations. 
In this Appendix, we present a cleaned catalog of GCs in NGC~5128.  We also provide additional notes on objects that we have ``demoted".  

It has been nearly a decade since an up-to-date comprehensive catalog of confirmed GCs around NGC~5128 has been published.  Recent papers have either studied a subset of candidates and known GCs or have relied on ``private communication" as one of their data sources.  With decades of observations and a multitude of naming conventions and selection criteria, we hope to aid in future studies by providing information about all confirmed GCs known at this time in a clear way.

\section{Table of Confirmed GCs} \label{app:confirmed}

We present a compilation catalog of the 557 confirmed GCs around NGC~5128.  
The most common way to confirm that a candidate GC is associated with NGC~5128 is to measure its radial velocity, since most GCs at that distance are too small to visually resolve into component stars with ground-based data.  NGC~5128 has a systematic velocity of 541 km s$^{-1}$  \citep{Hui1995}, and a central stellar velocity dispersion of $\sim$150 km s$^{-1}$ \citep{Wilkinson1986}.  Therefore, to avoid contamination from Milky Way foreground stars, past surveys have considered a GC to be confirmed if its radial velocity is greater than 250 km s$^{-1}$.
Three studies confirm select GCs based on morphological parameters if their radial velocity measurement is below this critical value \citep{Woodley2010b, Woodley2010a, Mouhcine2010}.  
Additionally, \citet{Harris2002, Harris2006} list new GCs found via their resolved stellar outskirts in {\it HST} STIS and ACS imaging, respectively, and some of these GCs do not currently have radial velocity measurements.  We consider all of these ``confirmed" GCs of NGC~5128, unless there exists evidence to the contrary.  

For each confirmed GC around NGC~5128, Table \ref{table:confirmed_short} first lists the ID from the GC candidate catalog in this paper, the discovery ID, and all references in the literature.  
Also listed are the right ascension and declination in degrees (J2000) as well as the $g$ and $r$ band magnitudes. These measurements are from the PISCeS dataset to provide uniform astrometric and photometric systems, and the magnitudes have not been extinction corrected. 
Next we include the most recently published radial velocity and error (km s$^{-1}$),  along with its reference.
In the final column, we note which GCs are in our fiducial sample (\S \ref{sec:test sample}) and two GCs that have slightly incorrect positions in followup surveys.  For completeness, if a GC is not in PISCeS or is within the 10' from the center of NGC~5128, we still include it in this catalog using its most recent astrometry and without PISCeS photometry.  



\begin{table*}[ht]
\tiny
\centering
\caption{Confirmed GCs around CenA}
\begin{tabular}{c c c c c c c c c c c}
\hline \hline
ID & Discovery ID & references* & R.A. & Decl. & radial distance  & $g$  &  $r$  & v$_r$         & v$_r$ Reference & Notes  \\
(H21) &           &        & (J2000) & (J2000) &  (arcminutes)  & (mag) & (mag)  & (km s$^{-1}$) &                &    \\
\hline
H21-069900 & Fluffy & z & 199.545349 & -44.157249 & 103.8 & 19.42 & 18.73 & 719 $\pm$ 6 & z &   \\
H21-154980 & KK197-01 & p;aa & 200.499166 & -42.535139 & 48.37 &       &       & 636 $\pm$ 16 & aa &   \\
H21-156130 & KK197-03 & p;aa & 200.510634 & -42.536913 & 47.90 & 21.41 & 20.75 & 642 $\pm$ 3 & aa &   \\
H21-159794 & H12-556 & w;y;v & 200.552063 & -42.753072 & 39.41 & 21.29 & 20.63 &   &   &   \\
H21-160341 & KKs55-02 & ab & 200.557787 & -42.734733 & 39.66 &       &       & 531 $\pm$ 15 & ab &   \\
H21-183480 & KV19-212 & z & 200.790878 & -43.874443 & 56.25 & 18.28 & 17.53 & 535 $\pm$ 2 & z & 1 \\
H21-194226 & 32 & i;k;n;o;y & 200.909700 & -42.773027 & 25.31 & 18.86 & 18.22 & 497 $\pm$ 33 & o & 1 \\
H21-195812 & C40 & f;g;i;k;n;o;y & 200.926424 & -43.160459 & 20.56 & 19.37 & 18.64 & 443 $\pm$ 48 & o &   \\
H21-196535 & VHH81-01 & b;c;g;i;n;o;r;s;x;y & 200.934075 & -43.186591 & 20.89 & 18.12 & 17.38 & 645 $\pm$ 10 & s &  3 \\
H21-196891 & GC0563 & r;y & 200.937603 & -43.019830 & 18.65 & 19.67 & 18.85 & 533 $\pm$ 35 & r & 1 \\

\hline
\end{tabular}
\begin{tablenotes}
      \small
      \item The complete table will be available online.
      \item *See Table \ref{table:cands_descr} for annotation references
      \item Notes: 1) GC is in our fiducial sample (see Section \ref{sec:test sample}); 
      
      \item 2) The position of GC pff-gc-028 is originally listed in \citet{Peng2004} in HMS coordinates as 13:25:01.16, $-$42:56:51.05.  In their Table 1, \citet{Woodley2007} gives the position of this GC as 13:20:01.16, --42:56:51.05, which due to an unfortunate change in one digit of right ascension minutes, points to an empty patch of sky.
      
      \item 3) The position of C01/GC0005 is listed in \citet{Taylor2010} as 200.93333, --43.187083, which is 2.65" away from the position indicated by all other studies that noted this GC. 
    
\end{tablenotes}

\label{table:confirmed_short}
\end{table*}




\section{Mis-Classified Objects} \label{app:misconfirmed}

In building a catalog of known GCs around NGC~5128, it is important to note instances of 73 targets that have been erroneously confirmed as GCs.  We find 50 targets that we reclassify as foreground stars based on {\it Gaia} DR2 proper motion and/or parallax measurements.  These are discussed in Section \ref{app:hvfs} and listed in Table \ref{table:hvfs}.  Additionally, there are also 32 occurrences when followup measurements of velocity, spectra, or structure do not agree with a target's initial classification (of which an additional 23 are objects that have been reclassified as foreground stars).  These are discussed in Section \ref{app:conflicting} and listed in Table \ref{table:conflicting}.  

There is one additional case in which the position listed for an identified GC points to an empty patch of sky that cannot be readily matched with a nearby known or candidate GC.  The position for GC0472 from \citet{Woodley2010b} 
is given as $\alpha$=201.42908$^{\circ}$, $\delta$=$-$43.00056$^{\circ}$. 
This is only 3.6\arcmin from the center of NGC~5128, a region crowded with stars and dust.  The position could refer to a source at $\alpha$=201.42822$^{\circ}$, $\delta$=$-$43.00056$^{\circ}$ (a distance of 2.3\arcsec away), but we do not want to make that assumption here, as no other NGC~5128 GC surveys have published observations of this object.  We note that there is a very good match between the astrometry used in \citet{Woodley2010b} and the PISCeS data used here, and so this mismatch is unique.

\subsection{Foreground stars with measured proper motions in {\it Gaia} DR2} \label{app:hvfs}

With the modest systematic velocity and velocity dispersion of NGC~5128, there is some overlap in the radial velocity measurements of NGC~5128 GCs and Milky Way foreground stars: a radial velocity measurement greater than 250 km s$^{-1}$ \emph{alone} cannot definitively give membership for every object.
Of the $\simeq$ 600 identified GCs in NGC~5128, 355 can also be found in {\it Gaia} DR2, and 50 of these have recorded proper motions and/or parallax with significance greater than 3$\sigma$.  
Their velocities are plotted as the blue histogram in Figure \ref{fig:rv_hist}. Because these objects have high-confidence apparent motion measurements, we can infer that they are Galactic foreground stars, rather than GCs in NGC~5128.  
These foreground stars can also be seen as distinct from the fiducial sample GCs in AEN--magnitude space and $BR_{\rm excess}$--magnitude space, as seen by the blue diamonds in Figure \ref{fig:aen_bprp}, showing that all of these objects match better to Gaia's standard astrometric model of a star than our fiducial sample of confirmed GCs do.
Five of these stars also have HST imaging and do not have the characteristic resolved stellar outskirts of true GCs.  
The stars with $\gtrsim$ 500 km s$^{-1}$ radial velocities are possibly being ejected from the Milky Way, although the uncertainties on many of these measurements are relatively large ($\approx$ 100 -- 200 km s$^{-1}$) and so warrant further observations.

For each foreground star that was previously thought to be a GC in NGC~5128, Table \ref{table:hvfs} first lists the right ascension and declination in degrees (J2000). The position measurements are all from the PISCeS dataset to provide a uniform astrometric system. 
Next we list the most recent reported radial velocity and error (km s$^{-1}$) along with its reference.  Columns 5--7 list the right ascension proper motion and error (mas yr$^{-1}$), the declination proper motion and error (mas yr$^{-1}$), and  the parallax motion and error (mas yr$^{-1}$) from {\it Gaia} DR2.  Finally we include the various names each has had in the literature, along with references.  In the list of names we give preference to the earliest naming convention when small differences are found, such as the removal or addition of preceding ``0" digits or the removal, addition, or slight alteration of prefixes.  

We identified two authentic GCs that have high apparent motion measurements in {\it Gaia} DR2 (see Notes 10 and 18 in Table \ref{table:conflicting}).  Because {\it Gaia} DR2 is designed to observe stars in the Milky Way, it does not always provide accurate measurements of sources that are extended or in very crowded regions.  We urge caution and follow-up observations of the objects in Table \ref{table:hvfs}, especially those within 10\arcmin of the center of NGC~5128.




\begin{table*}[ht]
\centering
\tiny
\caption{Foreground stars that were previously mis-classified as GCs}
\begin{tabular}{c c c c c c c c c}
\hline \hline
RA & Dec & v$_r$ & v$_r$ Ref.* & $\mu_{\alpha}$  & $\mu_{\delta}$ & $\varpi$ & Names* & Notes  \\
(deg J2000) & (deg J2000)& (km/s) &   & (mas/yr) & (mas/yr) & (mas/yr) &  & \\
\hline
200.858796 & -42.924206 & 235 $\pm$ 43 & r & -4.81$ \pm $0.42 & -0.79$ \pm $0.35 & 0.73$ \pm $0.26 & GC0560$^r$; T17-0955$^y$ & \\
200.867666 & -42.795083 & 181 $\pm$ 68 & r & -8.67$ \pm $0.16 & -0.08$ \pm $0.13 & 0.27$ \pm $0.09 & GC0561$^r$; T17-0962$^y$ & \\
200.868472 & -42.886790 & 238 $\pm$ 26 & r & -4.86$ \pm $0.49 & -3.69$ \pm $0.40 & 0.44$ \pm $0.30 & GC0562$^r$; T17-0964$^y$ & \\
200.977151 & -43.333608 & 274 $\pm$ 49 & r & -8.72$ \pm $1.59 & 0.51$ \pm $1.14 & -0.10$ \pm $0.74 & GC0007$^{n,r}$; AAT301956$^o$; T17-1035$^y$ & \\
200.994011 & -42.954703 & 590 $\pm$ 144 & r & -11.7$ \pm $0.99 & -0.34$ \pm $0.74 & 0.44$ \pm $0.57 & GC0009$^{n,r}$; AAT101931$^o$; T17-1048$^y$ & \\
200.998369 & -42.922017 & 293 $\pm$ 83 & o & -3.35$ \pm $1.03 & 0.28$ \pm $0.82 & -1.03$ \pm $0.61 & GC0012$^n$; AAT102120$^o$; T17-1054$^y$ & \\
201.024844 & -43.065200 & 277 $\pm$ 67 & o & -6.15$ \pm $1.09 & -8.05$ \pm $0.86 & -0.24$ \pm $0.72 & GC0016$^n$; AAT103195$^o$; T17-1068$^y$ & \\
201.027472 & -42.882931 & 285 $\pm$ 29 & r & -9.81$ \pm $0.35 & -2.19$ \pm $0.31 & 0.19$ \pm $0.23 & GC0565$^r$; T17-1071$^y$ & \\
201.035905 & -43.274189 & 305 $\pm$ 56 & n & 6.09$ \pm $1.01 & -5.08$ \pm $0.73 & 0.72$ \pm $0.63 & GC0017$^n$; T17-1081$^y$ & \\
201.040767 & -42.748422 & 173 $\pm$ 117 & r & -6.87$ \pm $0.46 & -2.66$ \pm $0.30 & 0.49$ \pm $0.21 & GC0566$^r$; T17-1084$^y$ & \\
201.137766 & -43.312476 & 344 $\pm$ 58 & n & -2.96$ \pm $0.96 & -4.63$ \pm $0.83 & 1.60$ \pm $0.56 & pff-gc-010$^{g,i,l}$; GC0035$^n$; T17-1180$^y$ & 1 \\
201.181460 & -43.145317 & 465 $\pm$ 38 & o & -4.66$ \pm $0.52 & -1.55$ \pm $0.43 & -0.13$ \pm $0.32 & GC0047$^{n,u}$; AAT109380$^o$; T17-1216$^y$ & \\
201.199229 & -43.145415 & 353 $\pm$ 29 & r & -3.25$ \pm $1.40 & -3.98$ \pm $1.14 & -1.07$ \pm $0.88 & GC0575$^{r,u}$;  T17-1240$^y$ & \\
201.216990 & -43.075755 & 195 $\pm$ 29 & q & -6.81$ \pm $1.24 & -2.84$ \pm $0.88 & -0.19$ \pm $0.64 & C100$^{f,q}$; GC0068$^{n,r,u}$; T17-1269$^y$ & \\
201.220601 & -43.198898 & 302 $\pm$ 166 & o & -10.1$ \pm $1.39 & -6.18$ \pm $1.23 & 0.07$ \pm $0.92 & GC0069$^n$; AAT111033$^o$; T17-1270$^y$ & \\
201.224941 & -43.073484 & 466 $\pm$ 87 & o & -5.11$ \pm $1.08 & -7.67$ \pm $0.84 & -0.78$ \pm $0.59 & GC0071$^{n,u,x}$; AAT111185$^o$;  T17-1278$^y$ & \\
201.228369 & -43.147004 & 197 $\pm$ 29 & r & -12.9$ \pm $0.71 & -3.60$ \pm $0.58 & 0.74$ \pm $0.43 & GC0577$^{r,u}$; T17-1289$^y$ & \\
201.235562 & -42.658724 & 158 $\pm$ 50 & r & 0.14$ \pm $0.36 & -2.75$ \pm $0.28 & 0.78$ \pm $0.16 & GC0579$^r$; T17-1306$^y$ & \\
201.237909 & -42.648595 & 170 $\pm$ 45 & r & -1.41$ \pm $0.54 & -1.61$ \pm $0.40 & 0.08$ \pm $0.23 & GC0580$^r$; T17-1311$^y$ & \\
201.269183 & -43.122730 & 456 $\pm$ 118 & o & 1.89$ \pm $0.78 & -4.51$ \pm $0.72 & 0.67$ \pm $0.47 & GC0118$^{n,u}$; AAT112964$^o$; T17-1385$^y$ & \\
201.287915 & -42.400245 & 456 $\pm$ 52 & o & -3.57$ \pm $0.52 & -3.26$ \pm $0.57 & 0.03$ \pm $0.28 & pff-gc-039$^{g,i,o}$; GC0133$^n$; T17-1422$^y$ & \\
201.292702 & -42.919294 & 576 $\pm$ 12 & r & 1.11$ \pm $1.14 & -1.43$ \pm $1.02 & -2.24$ \pm $0.69 & GC0137$^{n,r,u}$; K-033$^{o,q}$; T17-1429$^y$ & \\
201.296247 & -42.967546 & 330 $\pm$ 61 & o & 0.09$ \pm $0.62 & -14.0$ \pm $0.48 & 0.18$ \pm $0.32 & HGHH-G348$^{d,g,i,o}$; GC0142$^{n,u}$; T17-1437$^y$ & \\
201.308075 & -42.961843 & 373 $\pm$ 36 & o & -8.33$ \pm $0.51 & -0.37$ \pm $0.42 & 0.09$ \pm $0.27 & HGHH-G271$^{d,g,i,j,o}$; GC0155$^{n,u}$; T17-1464$^y$ & \\
201.311756 & -43.686271 & 502 $\pm$ 53 & o & -4.20$ \pm $0.65 & 1.67$ \pm $0.49 & 0.65$ \pm $0.31 & pff-gc-046$^{g,i,o}$; GC0159$^n$; T17-1472$^y$ & \\
201.318460 & -43.059162 & 352 $\pm$ 136 & n & -8.97$ \pm $1.81 & -2.93$ \pm $2.38 & -0.01$ \pm $1.07 & AAT114993$^l$; GC0170$^{n,u}$; T17-1487$^y$ & 1 \\
201.329612 & -43.201107 & 453 $\pm$ 232 & o & -18.9$ \pm $1.11 & -1.72$ \pm $0.93 & 0.56$ \pm $0.61 & GC0183$^n$; AAT320656$^o$; T17-1515$^y$ & \\
201.339502 & -43.324865 & 163 $\pm$ 57 & r & -25.6$ \pm $1.14 & -2.48$ \pm $0.85 & 0.62$ \pm $0.72 & GC0427$^r$; T17-1533$^y$ & \\
201.357052 & -42.627947 & 404 $\pm$ 74 & o & -16.7$ \pm $1.33 & 2.30$ \pm $1.14 & 1.72$ \pm $0.53 & GC0198$^n$; AAT204119$^o$; T17-1557$^y$ & \\

201.358160 & -43.057139 & 269 $\pm$ 27 & q & -8.33$ \pm $0.73 & -0.81$ \pm $0.73 & 0.40$ \pm $0.42 & HGHH-46$^{d,g,i,o}$; GC0200$^{n,u}$; GC0462$^q$; T17-1560$^y$; T17-1561$^y$ & 2 \\

201.358752 & -43.189572 & 243 $\pm$ 19 & r & -5.34$ \pm $0.41 & -2.25$ \pm $0.38 & -0.19$ \pm $0.27 & GC0519$^r$; T17-1562$^y$ & \\
201.370066 & -43.072667 & 553 $\pm$ 155 & r & -4.23$ \pm $0.32 & -2.76$ \pm $0.31 & 0.16$ \pm $0.20 & C145$^l$; GC0207$^{n,r,u}$; T17-1578$^y$ & 3 \\
201.376149 & -43.698210 & 297 $\pm$ 40 & n & -2.61$ \pm $1.64 & -5.08$ \pm $1.23 & -0.96$ \pm $0.77 & pff-gc-054$^g$; GC0216$^n$; T17-1593$^y$ & \\
201.394333 & -43.054556 & 585 $\pm$ 97 & r & -6.55$ \pm $0.29 & 0.44$ \pm $0.27 & 0.31$ \pm $0.18 & C152$^l$; GC0241$^{n,r}$; T17-1634$^y$ & 3 \\
201.396764 & -43.200422 & 243 $\pm$ 61 & n & -12.3$ \pm $0.73 & 1.40$ \pm $0.57 & 0.27$ \pm $0.33 & 58$^i$; WHH-21$^j$; GC0247$^n$; T17-1644$^y$ & \\


201.408318 & -43.283137 & 257 $\pm$ 59 & o & -15.1$ \pm $1.07 & -5.93$ \pm $1.07 & -0.98$ \pm $0.75 & GC0256$^n$; AAT118314$^o$; T17-1665$^y$ & \\
201.410147 & -43.084056 & 28 $\pm$ 9 & r & -4.19$ \pm $0.36 & -0.96$ \pm $0.35 & 0.59$ \pm $0.24 & C156$^l$; GC0258$^{n,r,u}$;  T17-1668$^y$ & 3 \\
201.434215 & -42.983167 & 550 $\pm$ 34 & q & -5.57$ \pm $1.71 & 1.58$ \pm $1.49 & 1.05$ \pm $1.05 & HHH86-35$^{c,g,i,q}$; GC0286$^{n,u}$; T17-1724$^y$ & \\
201.443677 & -42.581676 & 293 $\pm$ 115 & o & -8.23$ \pm $0.76 & -0.76$ \pm $1.05 & -0.05$ \pm $0.44 & GC0294$^n$; AAT208206$^o$; T17-1745$^y$ & \\
201.466443 & -43.204605 & 187 $\pm$ 41 & r & -9.60$ \pm $0.50 & -0.72$ \pm $0.49 & -0.18$ \pm $0.26 & GC0592$^r$; T17-1797$^y$ & \\
201.519553 & -42.793058 & 336 $\pm$ 160 & o & -0.49$ \pm $1.32 & -5.68$ \pm $1.46 & 0.46$ \pm $0.66 & GC0363$^n$; AAT122794$^o$; T17-1914$^y$ & \\
201.550553 & -42.817643 & 225 $\pm$ 35 & r & -2.88$ \pm $0.70 & -5.63$ \pm $1.02 & 0.82$ \pm $0.38 & GC0443$^r$; T17-1964$^y$ \\
201.566173 & -42.916896 & 582 $\pm$ 10 & s & -2.83$ \pm $0.74 & -2.05$ \pm $0.88 & -0.33$ \pm $0.42 & f1.GC-22$^i$; R122$^m$; GC0382$^{n,r,s,x}$; T17-1974$^y$ \\
201.599416 & -42.788083 & 343 $\pm$ 43 & n & -10.9$ \pm $0.15 & -4.85$ \pm $0.15 & 2.26$ \pm $0.07 & HGHH-51$^g$; GC0398$^n$; T17-2015$^y$ \\
201.599715 & -43.295681 & 277 $\pm$ 158 & o & -6.07$ \pm $1.47 & -1.01$ \pm $2.46 & -1.06$ \pm $0.84 & GC0399$^n$; AAT335187$^o$; T17-2017$^y$ \\
201.657114 & -42.981575 & 276 $\pm$ 22 & r & -5.66$ \pm $0.54 & -13.0$ \pm $0.60 & -0.17$ \pm $0.31 & GC0553$^r$; T17-2063$^y$ \\
201.674875 & -43.129149 & 271 $\pm$ 20 & n & -50.3$ \pm $0.20 & -11.7$ \pm $0.28 & 0.92$ \pm $0.11 & HHH86-39$^{c,g,i,j,o}$; GC0406$^{n,u}$; T17-2078$^y$ \\
201.712004 & -43.122378 & 372 $\pm$ 42 & r & -10.1$ \pm $3.23 & 2.25$ \pm $4.16 & 2.13$ \pm $1.55 & GC0557$^{r,u}$; T17-2104$^y$ \\
201.839971 & -42.644833 & 252 $\pm$ 71 & o & 8.28$ \pm $0.49 & -9.16$ \pm $0.59 & 0.29$ \pm $0.25 & pff-gc-101$^{g,i,o}$; GC0414$^n$; T17-2180$^y$ \\
201.753258 & -42.360852 & 594 $\pm$ 102 & o & -10.7$ \pm $1.26 & -3.20$ \pm $1.30 & 0.17$ \pm $0.74 & AAT223403$^o$ \\

\hline
\end{tabular}
\begin{tablenotes}
      \small
      \item *See Table \ref{table:cands_descr} for annotation references
      \item Notes: 1)  This object was found to be a star based on \textit{Hubble Space Telescope (HST)}/Advanced Camera for Surveys (ACS) imaging by \citet{Harris2006}.
      \item 2) See Appendix \ref{app:too many names} for information about the naming history of this object.
      \item 3)  This object was identified as a GC based on  {\it HST} ACS
      imaging by \citet{Harris2006}.  It also has large ($\sigma \geq 4$) proper motion and/or parallax measurements in {\it Gaia} DR2, and careful re-inspection of its {\it HST} imaging confirms it to be a point source.  
\end{tablenotes}
\label{table:hvfs}
\end{table*}

\subsection{Conflicting measurements} \label{app:conflicting}

Thirty-two of the identified GCs in NGC~5128 have conflicting classifications in the literature.  A majority of these conflicting classifications occur because the target was measured to have a radial velocity within the standard acceptable range of $\approx$ 250 -- 1000 km s$^{-1}$ by one group, but outside this range by another.  Additionally, two objects were identified as GCs based on their radial velocities, but later found to be stars based on {\it HST}/Advanced Camera for Surveys (ACS) imaging (see note 9 in Table \ref{table:conflicting}). 
Another object 
was subsequently identified as a star because of its spectrum (see note 5 in Table \ref{table:conflicting}). 
These objects with conflicting measurements have been matched between studies based on position, and in many cases by their ID as well.  When examining these objects with PISCeS data, several of them can be visually identified as background galaxies. 
We believe seven of these objects are indeed true GCs, because a majority of their measurements classify them as such and they do not visually look like galaxies, so they are included in Table \ref{table:confirmed_short}. 

For each object with conflicting measurements, Table \ref{table:conflicting} first lists the right ascension and declination in degrees (J2000). The position measurements are from the PISCeS dataset to provide a uniform astrometric system. Next we list the conflicting radial velocities and errors (km s$^{-1}$) along with their sources, if applicable.  We also include our classification of each object, as well as further information indicated by the ``Notes" column.

Many sources experience a discrepancy between \citet{Woodley2007} and \citet{Beasley2008}.  It appears that between sharing data with the Woodley group and publishing their own paper, the Beasley group updated measurements on several targets.  
We note that one of these objects has a mismatch in both its velocity measurement and coordinates.
\citet{Woodley2007} gives an incorrect position for HH-048 as $\alpha$=200.68900$^{\circ}$, $\delta$=$-$43.119111$^{\circ}$, with a radial velocity of 856 $\pm$ 56 km s$^{-1}$, citing  \citet{Beasley2008} as the reference for this radial velocity measurement.  
\citet{Beasley2008}, however, lists HH-048 as a background galaxy ($v_r$ = 16000 km s$^{-1}$) with the same right ascension but a declination  of $\delta$=$-$43.011917$^{\circ}$ (note the single difference in the digit following the decimal point).  Based upon visual inspection, the position listed in the Beasley catalog is correct, and is listed in Table \ref{table:conflicting}

\begin{table*}[ht]
\tiny
\centering
\caption{Previously Classified Objects with Conflicting Measurements}
\begin{tabular}{c c c c c c c c }
\hline \hline
RA & Dec & v$_r$ \#1  & Source \#1* & v$_r$ \#2 & Source \#2* & Our Classification & Notes   \\ 
(deg J2000) & (deg J2000)& (km/s)  &  & (km/s)&   &  & \\
\hline

200.689000  & -43.011917 &  856 $\pm$ 56 & m & 16000 & n & galaxy & 1, 2 \\ 

200.926424 & -43.160459 & 357 $\pm$ 42 & f &     & j & GC & 3 \\

200.986186 & -43.000083 & 798 $\pm$ 49 & m & 13000  & n & galaxy & 1 \\ 

201.011383 & -42.809021 & 775 $\pm$ 75 & m & 5000  & n & galaxy & 1 \\ 

201.035905 & -43.274189 &  305 $\pm$56 & m &  & n & star & 4, 5  \\ 

201.099937 & -42.902944 & 582 $\pm$ 81 & i & 46000 & n & unknown & 6 \\ 

201.117323 & -42.884562 & 835 $\pm$ 83 & m & 13000  & n & unknown & 7 \\ 

201.134014 & -43.182469 & 636 $\pm$ 78 & i &    & j &  GC & 8 \\ 

201.137766 & -43.312476 & 344 $\pm$ 58 & f & & k & star & 4, 9 \\ 

201.144203 & -43.214044 & 7717 $\pm$ 194 & i & 517 $\pm$ 123 & m & unknown &  \\ 

201.188363 & -43.388845 & 998 $\pm$ 250 & m & 17000 & n & galaxy & 1 \\ 

201.203346 & -43.271446 & 42000  & n & 426 $\pm$ 41 & r & unknown & \\ 

201.232175 & -43.379980 & 921 $\pm$ 146 & m & 9000  & n & galaxy & 1 \\ 

201.263696 & -43.137316 & 31000  & n & 690 $\pm$ 31 & r & unknown &  \\ 

201.318460 & -43.059162 & & k & 352 $\pm$ 136 & n & star & 4, 9 \\ 

201.347769 & -42.890588 & 2017 $\pm$ 78 & i & 545 $\pm$ 64 & m & unknown &   \\ 


201.376149 &  -43.698210 & 317 $\pm$ 43 & f & 169 $\pm$ 108 & n & star & 4 \\   


201.406112 & -43.095806 & & l & & this work & GC & 10 \\


201.415196 & -43.067039 & 510 $\pm$ 33 & f & 32124 $\pm$ 253 & i & GC & 11 \\ 

201.440728 & -42.571614 & 998 $\pm$ 250 & m & 15000 & n & galaxy & 1 \\ 

201.455161 & -43.000611 & 7349 $\pm$ 87 & i & 465 $\pm$ 30 & q & unknown & 12 \\ 

201.455309 & -43.039000 & 384 $\pm$ 30 & f & 9942 $\pm$ 113 & i & GC & 13 \\ 

201.543947 & -43.018297 & 231 $\pm$ 143 & n & 311 $\pm$ 81 & r & unknown & \\ 

201.550553 & -42.817643 & 117 $\pm$ 80 & n & 225 $\pm$ 35 & r & star & 4, 14 \\ 

201.572047 & -43.110904 & 24742 $\pm$ 79 & i & 513 $\pm$ 183 & m & unknown &  \\ 

201.574090 & -42.770562 & 64000 & n & 178 $\pm$ 51 & r & galaxy & 1 \\ 

201.585963 & -42.896082 & 271 $\pm$ 64 & r &  & this work & unknown &  15 \\

201.592072 & -43.152950 & 505 $\pm$ 78 & i & 21000 & n & unknown & 16 \\ 

201.660454 & -42.762675 & 474 $\pm$ 65 & f & 2000 & n & GC & 17 \\ 

201.705431 & -43.082693 &  839 $\pm$ 63 & m & 15000 & n & galaxy & 1 \\ 

201.748983 & -42.924119 & 833 $\pm$ 32 & m & 13000 & n & galaxy & 1 \\ 

202.076936 & -42.553438 & 410 $\pm$ 42 & f &      & this work & GC & 18 \\

\hline
\end{tabular}
\begin{tablenotes}
      \footnotesize
      \item *See Table \ref{table:cands_descr} for annotation references

      \item Notes: 1) Based on visual inspection with PISCeS data, this object is a galaxy.
      
      \item 2) See discussion in Section~\ref{app:conflicting} about the position of this object in the literature.
      
      \item 3) This object also has a radial velocity measurement of 443 $\pm$ 48 km s$^{-1}$ from \citet{Beasley2008}, no parallax or proper motion and a high $\sigma$ AEN value in {\it Gaia} DR2.   We believe that the previous classification of this object as a foreground star is inaccurate (based on the shape of its PSF subtracted residual by \citet{Gomez2006}), and so include it in our confirmed GC sample.

      \item 4) Based on {\it Gaia} DR2 proper motion and parallax data, we find this object to be foreground star.  It is also listed in Table \ref{table:hvfs}.
      
      \item 5) This object was identified as a star by \citet{Beasley2008} because it exhibits strong molecular bands in its spectrum that are characteristic of an M-type star.
      
      \item 6) This object is also measured with $v_r$ = 606 $\pm$ 33 km s$^{-1}$ in \citet{Woodley2010a}, but they still list the older velocity measurement of $v_r$ = 582 $\pm$ 81 as their ``new weighted radial velocity".  It was also noted as a background galaxy in \citet{Gomez2006} based on the shape of its PSF subtracted residual.
      
      \item 7) Based on visual inspection with PISCeS data, at this position there is a galaxy and an unknown, point-source-like object projected onto each other. 
      
      \item 8) This object also has a radial velocity measurement of 775 $\pm$ 74 from \citet{Beasley2008} and 813 $\pm$ 40 from \citet{Woodley2010a}, and no parallax or proper motion and a high $\sigma$ AEN value in {\it Gaia} DR2.   We believe the classification as a background galaxy by \citet{Gomez2006} based on the shape of its PSF subtracted residual is inaccurate, and so include it in our confirmed GC sample.

      \item 9) This object was found to be a star based on {\it HST} ACS imaging by \citet{Harris2006}.

      
      \item 10) This object was initially identified as a GC based on {\it HST} ACS 
      imaging by \citet{Harris2006}, and it has two radial velocity measurements of 575 $\pm$ 36 \citep{Woodley2007} and 524 $\pm$ 18 \citep{Woodley2010b}.  In {\it Gaia} DR2, it has a proper motion measurement with $\sigma$ = 4.12, but in {\it Gaia} EDR3 it has a proper motion measurement with only $\sigma$ = 1.5.  Because it is very close to the center of the galaxy, we believe the {\it Gaia} DR2 measurements to be inaccurate, and so include it in our confirmed GC sample.  
      
      \item 11) This was also measured by \citet{Beasley2008} with $v_r$ = 557 $\pm$ 49 km/s.  We believe the classification as a galaxy is inaccurate, and so include it in our confirmed GC sample. 
      
      \item 12) This object is  also measured with $v_r = 475 \pm 134$ in \citet{Woodley2010a}, but they still list the older velocity measurement of $v_r$ = 465 $\pm$ 30 as their ``new weighted radial velocity".
      
      \item 13) This object was also measured by \citet{Beasley2008} to have $v_r = 420 \pm 51$, and by \citet{Woodley2010a} to have $v_r$ = 371~$\pm$~28.  We believe the classification as a galaxy is inaccurate, and so include it in our confirmed GC sample. 
      
      \item 14) \citet{Woodley2010a} classified a small number of candidates as GCs with low velocities based on their structural parameters, many of which we find to actually be foreground stars.  
      
      \item 15) This object has one radial velocity measurement of 271 $\pm$ 64 km s$^{-1}$, but also has stellar AEN and $C_{3-6}$ values.  It has no reported proper motion or parallax in {\it Gaia} DR2. We recommend that further study of this GC to more confidently reinforce or oppose its status as a GC of NGC~5128.
      
      \item 16)  This object is also measured with $v_r$ = 357 $\pm$ 35 km s$^{-1}$ in \citet{Woodley2010a}, but they still list the older velocity measurement of $v_r$ = $505 \pm$ 78 as their ``new weighted radial velocity".
      
      \item 17) The position of this target is listed by \citet{Peng2004, Beasley2008} as 201.65829, $-$42.763861, which is ~7" away from the position we, and other followup studies, find for this GC.  It also has additional radial velocity measurements of 492 $\pm$ 37 km s$^{-1}$ in \citet{Woodley2007} and 627 $\pm$ 21 km s$^{-1}$ in \citet{Woodley2010a}.
      
      \item 18) This object has a {\it Gaia} DR2 proper motion measurement of 13.47 $\pm$ 2.68 mas yr$^{-1}$.  We do not include it in the HVFS table, however, because it has AEN, parallax, color, and $C_{3-6}$ values and two independent radial velocity measurements consistent with its status as a GC in NGC~5128.  Therefore, it is is recorded in Table \ref{table:confirmed_short}, but is not included in the fiducial sample, Figure \ref{fig:rv_hist}, or Table \ref{table:hvfs}.

    \end{tablenotes}
\label{table:conflicting}
\end{table*}

\section{An object with too many names} \label{app:too many names}

During our literature review, we found an instance where two confirmed GCs with different names shared the same position. The objects identified most recently as  GC0462/T17-1560 and GC0200/T17-1561 have the same position in the \citet{Sinnott2010} catalog, and positions that differ by 0.24" in the T17 catalog.  Upon visual inspection, both objects seem to correspond to the GC originally identified as HGHH-46 in the \citet{Harris1992} catalog. The discrepancy appears to have begun when the GC was reported in the ``New Radial Velocity Measurements of Known Globular Clusters" table 
in \citet{Woodley2010b} even though it had already been identified by previous surveys.  Based on the object's high proper motion in {\it Gaia} DR2, we have determined that it is actually a foreground star rather than a GC in NGC~5128; see Appendix \ref{app:hvfs} for details.  We list all the names for this object in Table \ref{table:hvfs} in the hope of avoiding future confusion.

\section{Background galaxy positions that point to empty sky}\label{app:gal bad pos}
When compiling our list of radial velocity confirmed background galaxies, we found three instances of positions pointing to empty sky in the PISCeS dataset rather than an object, based upon visual inspection.  
These were originally noted in \citet{Harris2004} as new GC candidates, and confirmed as background galaxies by \citet{Woodley2005} based on radial velocity measurements.  Both of these papers generally have consistent astrometry with PISCeS data except for these instances.  The three positions in question are: (1) $\alpha$=201.01875$^{\circ}$, $\delta$=$-$42.804500$^{\circ}$, ID: 283; (2) $\alpha$=201.02288,$^{\circ}$, $\delta$=$-$42.790028$^{\circ}$, ID: 285; and (3) $\alpha$=201.25537,$^{\circ}$, $\delta$=$-$42.907361$^{\circ}$, ID: 299, where all positions are in J2000 coordinates.  We do not include these objects when calculating the fidelity of our extended galaxy checks in Section \ref{sec:galcul}.

\end{document}